\documentclass[aps,prc,preprint,amsmath,showpacs,superscriptaddress,nofootinbib]{revtex4-1}
\usepackage{graphicx,epstopdf,epsfig,psfrag}
\newcommand{\beqy}{\begin{eqnarray}}
\newcommand{\eeqy}{\end{eqnarray}}
\newcommand{\bmlet}{\begin{subequations}}
\newcommand{\emlet}{\end{subequations}}

\begin{document}

\title{Binary and ternary ionic compounds in the outer crust of a cold nonaccreting neutron star}

\author{N. Chamel}
\affiliation{Institut d'Astronomie et d'Astrophysique, CP-226, Universit\'e Libre de Bruxelles, 
1050 Brussels, Belgium}
\author{A. F. Fantina}
\affiliation{Institut d'Astronomie et d'Astrophysique, CP-226, Universit\'e Libre de Bruxelles, 
1050 Brussels, Belgium}
\affiliation{Grand Acc\'el\'erateur National d'Ions Lourds (GANIL), CEA/DRF - CNRS/IN2P3, Bvd Henri Becquerel, 14076 Caen, France}

\begin{abstract}
The outer crust of a cold nonaccreting neutron star has been generally assumed to be stratified into different layers, each of which 
consists of a pure body-centered cubic ionic crystal in a charge compensating background of highly degenerate electrons. The validity of this 
assumption is examined by analyzing the stability of multinary ionic compounds in dense stellar matter. It is thus shown that their stability  
against phase separation is uniquely determined by their structure and their composition irrespective of the stellar conditions. However, 
equilibrium with respect to weak and strong nuclear processes imposes very stringent constraints on the composition of multinary compounds, and 
thereby on their formation. By examining different cubic and noncubic lattices, it is found that substitutional compounds having the same 
structure as cesium chloride are the most likely to exist in the outer crust of a nonaccreting neutron star. The presence of ternary compounds is 
also investigated. Very accurate analytical expressions are obtained for the threshold pressure, as well as for the densities of the different 
phases irrespective of the degree of relativity of the electron gas. Finally, numerical calculations of the ground-state structure and of the 
equation of state of the outer crust of a cold nonaccreting neutron star are carried out using recent experimental and microscopic nuclear mass tables. 
\end{abstract}


\maketitle

\section{Introduction}

Neutron stars are formed in the aftermath of gravitational core-collapse of single massive stars with a mass $M > 8 M_\odot$, $M_\odot$ being 
the mass of the Sun~\cite{hae07}. During the collapse and the subsequent cooling of the hot compact stellar remnant, the compressed stellar material is generally 
assumed to follow a sequence of full thermodynamic quasi equilibrium states such that the resulting neutron star eventually consists of ``cold catalyzed matter'', 
i.e., electrically charge neutral matter in its absolute ground state at temperature $T=0$~K~\cite{hw58,htww65}. 

Under these assumptions, the outermost region of a neutron star is generally thought to form a solid crust stratified into different layers, each 
of which consists of a perfect crystal made of a single nuclear species with atomic number $Z$ and mass number $A$ (see, e.g., Ref.~\cite{lrr} and 
references therein). Due to the huge gravitational pressure, the density increases sharply with depth below the stellar surface. As the density reaches 
$\rho_{\rm eip}\simeq 11 AZ $~g~cm$^{-3}$, atoms are so densely packed that their electron clouds overlap (see, e.g., Ref.~\cite{hae07}). At densities 
$\rho\gg \rho_{\rm eip}$, atoms are thus fully ionized, and each crustal layer can be treated to a good approximation as a one-component crystal of 
pointlike ions (nuclei) in a uniform charge compensating background of highly degenerate electrons. It has been generally assumed that nuclei are arranged in 
a body-centered cubic (bcc) lattice, as put forward by Ruderman~\cite{ruderman68} based on the pioneer cubic-lattice constant calculations of 
Fuchs~\cite{fuchs35}. 

With increasing depth, matter becomes progressively more neutron rich due to the capture of electrons by nuclei (see, e.g., Ref.~\cite{chf15}). 
At density $\rho_{\rm drip}\simeq 4\times 10^{11}$~g~cm$^{-3}$, neutrons start to ``drip'' out of nuclei. The onset of neutron emission marks the 
transition between the outer and inner regions of the crust (see, e.g., Ref.~\cite{cfzh15} for a recent discussion). 
The outer crust can thus be described by a stack of pure bcc crystalline layers whose composition is completely determined by nuclear masses~\cite{bps71} 
(see, e.g., Refs.~\cite{roca2008,pearson2011,hemp2013,wolf13,chamel2015c,utama2016} for recent calculations). In 1971, Dyson~\cite{dyson71} suggested 
the existence of FeHe compound with rocksalt (NaCl) structure in the crust of a neutron star. This possibility was further studied by Witten in 1974~\cite{witten74}. 
However, as pointed out by Jog and Smith~\cite{jog82}, such a compound is unstable against weak and strong nuclear processes. On the other hand, they found that 
binary compounds with cesium chloride (CsCl) structure can be energetically favored at the interface between two adjacent crustal layers. 
More recent studies have focused on the formation of multinary ionic compounds in the crust of accreting neutron stars (see, e.g. Refs.~\cite{horo09,eng16}). The 
accretion of matter (mostly hydrogen and helium) from a companion star triggers a series of nuclear reactions, whose ashes sink deep into the crust 
(see, e.g. Ref.~\cite{lrr} for a review). This material eventually solidifies with its composition remaining essentially unchanged. 

In this paper, we pursue the investigation of the existence of ionic compounds in the outer crust of a cold nonaccreting neutron star. In Section~\ref{sec:transition}, 
we first examine the thermodynamic stability of a pure solid phase against the formation of a multinary compound with an arbitrary composition. The   
specific case of two-component solid phases is considered in Section~\ref{sec:transition2}, where the impact of solid-solid phase transitions 
on the equation of state of dense matter is discussed in detail. Using recent experimental and microscopic nuclear mass tables, we thereafter 
determine the ground-state structure of the outer crust of a neutron star allowing for ionic compounds. Results are presented and discussed in 
Section~\ref{sec:composition}.

\section{General thermodynamic considerations on phase transitions in cold dense matter}
\label{sec:transition}

In the following, we shall consider matter at temperatures $T$ below the crystallization temperature $T_m$ (for all practical purposes, we shall set $T=0$), 
and at densities $\rho$ above the ionization threshold and below the neutron-drip transition ($\rho_{\rm eip} \ll \rho \leq \rho_{\rm drip}$). 
It will be further assumed that all possible weak and strong nuclear reactions are allowed.

\subsection{Stability of a pure solid phase against a transition into a multi-component solid phase}
\label{sec:stability}

Let us examine the absolute stability of a solid made of only one type of nuclei $(A,Z)$ with mass number $A$ and atomic number $Z$
 at some pressure $P$ against the transition to a multi-component solid made of nuclei $(A_i,Z_i)$, where the index $i$ runs over the different nuclear species present. Let $n_i$ be the number density of nuclei $(A_i,Z_i)$. Their proportion $\xi_i$ is defined by 
\begin{equation}\label{eq:xi}
\xi_i=\frac{n_i}{\sum_j n_j}\, .
\end{equation}
Introducing the mean nucleon number density
\begin{equation}\label{eq:nbaryon}
\bar n=\sum _i A_i n_i \, ,
\end{equation}
 the nuclei number densities can be expressed as
\begin{equation}\label{eq:nion}
 n_i=\xi_i \frac{\bar n}{\bar A}\, , 
\end{equation}
where
\begin{equation}
 \bar A= \sum_i \xi_i A_i
\end{equation} 
denotes the mean mass number. 
The energy density $\mathcal{E}_N$ of nuclei is given by 
\begin{equation}
\label{eq:EN}
\mathcal{E}_N=\sum_i n_i M^\prime(A_i,Z_i)c^2 \, ,
\end{equation}
where $c$ is the speed of light and $M^\prime(A,Z)$ denotes the nuclear mass (including the rest mass of $Z$ protons, $A-Z$ neutrons and 
$Z$ electrons\footnote{The reason for including the electron rest mass is that experimental \emph{atomic} masses are generally 
tabulated rather than \emph{nuclear} masses.}). The nuclear mass $M^\prime(A,Z)$ can be obtained from the \emph{atomic} mass $M(A,Z)$ after 
subtracting out the binding energy of the atomic electrons (see Eq.~(A4) of Ref.~\cite{lpt03}). Ignoring the 
small quantum zero-point motion of ions about their equilibrium position, nuclei do not contribute to the pressure, i.e. $P_N=0$. 

Nuclei are embedded in a neutralizing electron background of number density $n_e$ given by 
\begin{equation}\label{eq:ne}
n_e=\sum_i Z_i n_i=y_e \bar n\, ,
\end{equation}
where $y_e=\bar Z/\bar A$ is the mean electron fraction defined in terms of the mean atomic number  
\begin{equation}
\bar Z=\sum_i \xi_i Z_i\, .
\end{equation}
Because electrons are highly degenerate, they can be very well described by an ideal relativistic Fermi gas 
(see, e.g., Ref.~\cite{chf16} for a discussion of the validity of this approximation). 
The expressions for the corresponding energy density $\mathcal{E}_e$ and pressure $P_e$ can be found in Chap. 2 of Ref.~\cite{hae07}. 
The main correction to the Fermi gas model arises from electron-ion interactions, which from 
dimensional analysis can be generally expressed as
\begin{equation}\label{eq:EL}
\mathcal{E}_L  = C e^2 n_e^{4/3} f(\{Z_i\})\, ,
\end{equation}
\begin{equation}
\label{eq:PL}
P_L=\frac{\mathcal{E}_L}{3}=\frac{C}{3} e^2 n_e^{4/3} f(\{Z_i\})\, ,
\end{equation}
respectively, where $e$ is the proton electric charge, while the structure constant $C < 0 $ and the dimensionless function $f(\{Z_i\})$ 
depend on the spatial arrangement of nuclei and on their charge only ($\{Z_i\}$ denotes the set of all charge numbers). The structure constant 
$C$ is normalized such that the structure function for solids made of isotopes ($Z_i=Z$ for all $i$) reduces to $f(Z)=Z^{2/3}$ (this includes 
the limiting case of a single-constituent phase). Note that Eqs.~(\ref{eq:EL}) and (\ref{eq:PL}) could also be applied to liquid phases with 
suitable values for the structure constant. 

The ground state of matter at pressure $P$ (and temperature $T=0$) is determined by the minimum of the Gibbs free energy per nucleon defined by 
\begin{equation}
 g=\frac{\mathcal{E}+P}{\bar n}\, , 
\end{equation}
where the mean energy density $\mathcal{E}$ of matter and the pressure $P$ are given by 
($m_e$ is the electron mass)
\begin{equation}
\label{eq:energy}
\mathcal{E}=\mathcal{E}_N + \mathcal{E}_e + \mathcal{E}_L  - n_e m_e c^2\, , 
\end{equation}
\begin{equation}
\label{eq:P}
 P=P_e+P_L\, , 
\end{equation}
respectively. 
The last term in Eq.~(\ref{eq:energy}) is introduced to avoid double counting. 
Collecting all terms using the thermodynamic identity $\mathcal{E}_e+P_e = n_e \mu_e$ ($\mu_e$ denoting the electron Fermi energy), the Gibbs free 
energy can be finally expressed as 
\begin{eqnarray}\label{eq:gibbs}
g(\{A_i,Z_i\},P)&=&\frac{\bar M^\prime c^2}{\bar A}+y_e\biggl[\mu_e(n_e) - m_e c^2  + \frac{4}{3} C e^2 n_e^{1/3} f(\{Z_i\}) \biggr]\, ,
\end{eqnarray}
where we have introduced the mean mass of nuclei
\begin{equation}
\bar M^\prime=\sum_i \xi_i M^\prime(A_i,Z_i) \, .
\end{equation}
The electron number density $n_e$ is related to the pressure $P$ through Eq.~(\ref{eq:P}). 

In its ground-state, a pure solid is expected to have a bcc structure. Although the absolute stability of this lattice still 
remains to be demonstrated, so far no other structure has been found to be more stable (see, e.g., Refs.~\cite{fuchs35,foldy78,bald92}). 
As a matter of fact, the value of the corresponding structure constant $C_{\rm bcc}$ (see, e.g. Table~\ref{tab1}) lies very close to the 
lower bound obtained by Lieb and Narnhofer~\cite{lieb75}
and corresponding to the ion-sphere model~\cite{salp54}, 
\begin{equation}
\label{eq:Clow}
C\geq C_{\rm min}\equiv -\frac{9}{10}\left(\frac{4\pi}{3}\right)^{1/3} \simeq -1.450793\, .
\end{equation}
From now on, we consider this conjecture to be true. 
For a multi-component solid to be more stable, the corresponding Gibbs free energy per nucleon must be lower. The threshold pressure $P_{\rm thres}$ 
for the onset of the transition can be obtained from the condition 
\begin{equation}\label{eq:gibbs-transition-general}
 g(A,Z,P_{\rm thres})=g(\{A_i,Z_i\},P_{\rm thres})\, .
\end{equation}
It is particularly convenient to rescale the function $f(\{Z_i\})$ as 
\begin{equation}
 \tilde{f}(\{Z_i\})\equiv (C/C_{\rm bcc})f(\{Z_i\})\, .
\end{equation}
Excluding situations such that $Z/A=\bar Z/\bar A$ (see Appendix~\ref{appendix1}) and 
expanding $g$ to first order in $\alpha=e^2/(\hbar c)$ ($\hbar$ is the Planck-Dirac constant), 
we find 
\begin{equation}\label{eq:threshold-condition}
% \mu_e + C_{\rm bcc}\, \alpha \hbar c n_e^{1/3}\biggl(\frac{4}{3}\frac{Z^{5/3}}{A} - \frac{1}{3}\frac{Z^{2/3} \bar Z}{\bar A} -\frac{\bar Z}{\bar A}\tilde{f}(Z_1,Z_2)\biggr)
% \left(\frac{Z}{A}-\frac{\bar Z}{\bar A}\right)^{-1} = \mu_e^{\rm thres}\, ,
\mu_e + C_{\rm bcc}\, \alpha \hbar c n_e^{1/3}F(Z,A ; \{Z_i, A_i\}) = \mu_e^{\rm thres}\, ,
\end{equation}
where we have introduced the threshold electron Fermi energy
\begin{equation}\label{eq:muethres}
\mu_e^{\rm thres}\equiv \biggl[\frac{\bar M^\prime c^2}{\bar A}-\frac{M^\prime(A,Z)c^2}{A}\biggr]\left(\frac{Z}{A}-\frac{\bar Z}{\bar A}\right)^{-1} +  m_e c^2\, ,
\end{equation}
and 
\begin{equation}\label{eq:def-F}
F(Z,A ; \{Z_i, A_i\})\equiv \left(\frac{4}{3}\frac{Z^{5/3}}{A} - \frac{1}{3}\frac{Z^{2/3} \bar Z}{\bar A} -\frac{\bar Z}{\bar A}\tilde{f}(\{Z_i\})\right)
\left(\frac{Z}{A}-\frac{\bar Z}{\bar A}\right)^{-1} \, .
\end{equation}
In Eq.~(\ref{eq:threshold-condition}), $n_e$ refers to the electron density of the pure solid of nuclei $(A,Z)$ at pressure $P_{\rm thres}$. The electron density of the two-component solid at the \emph{same} pressure is given by $n_e+\delta n_e$, where 
\begin{equation}\label{eq:deltane}
 \delta n_e \approx \frac{1}{3} C_{\rm bcc} e^2 n_e^{4/3} \biggl[Z^{2/3}-\tilde{f}(\{Z_i\})\biggr]\left(\frac{dP_e}{d n_e}\right)^{-1}\, . 
\end{equation}
Note that Eqs.~(\ref{eq:threshold-condition}) and (\ref{eq:deltane}) were obtained without making use of the actual expressions for the electron Fermi energy 
$\mu_e$ and pressure $P_e$. Therefore, these equations still remain valid in the presence of a strongly quantizing magnetic field as in the crust of magnetars 
(the lattice energy density $\mathcal{E}_L$ is independent of the magnetic field according to the Bohr-van Leeuwen theorem~\cite{bvl32}). 
In the absence of magnetic fields, Eq.~(\ref{eq:threshold-condition}) can be transformed into a quadratic polynomial equation and can thus be solved analytically, as demonstrated in the next section. 

\subsection{Transition pressure and densities of the solid phases}
\label{sec:threshold-conditions}

Recalling that the electron Fermi energy  is given by 
\begin{equation}
 \mu_e = m_e c^2 \sqrt{1+x_r^2}\, , 
\end{equation}
where  $x_r=\lambda_e k_e$ is a dimensionless relativity parameter, $\lambda_e=\hbar /(m_e c)$ is the electron Compton wavelength, 
and $k_e=(3\pi^2 n_e)^{1/3}$ is the electron Fermi wave number, the threshold condition~(\ref{eq:threshold-condition}) can be equivalently expressed as
\begin{equation}\label{eq:threshold-condition2}
x_r^2\left(1-\tilde F(Z,A ; \{Z_i, A_i\})^2\right) + 2 \gamma_e^{\rm thres} \tilde F(Z,A ; \{Z_i, A_i\}) x_r =  (\gamma_e^{\rm thres})^2-1\, , 
\end{equation}
with 
\begin{equation}\label{eq:gammae-thres}
\gamma_e^{\rm thres}\equiv \frac{\mu_e^{\rm thres}}{m_e c^2}\, , 
\end{equation}
\begin{equation}\label{eq:Ftilde}
\tilde F(Z,A ; \{Z_i, A_i\})\equiv \frac{C_{\rm bcc}}{(3\pi^2)^{1/3}}\alpha F(Z,A ; \{Z_i, A_i\})\, .
\end{equation}
Solving Eq.~(\ref{eq:threshold-condition2}) for $x_r$ yields   
\begin{eqnarray}\label{eq:exact-xr}
 x_r&=&\gamma_e^{\rm thres} \Biggl[\sqrt{1-\left(1-\tilde F(Z,A ; \{Z_i,A_i\})^2\right)/(\gamma_e^{\rm thres})^{2}}-\tilde F(Z,A ; \{Z_i, A_i\})\Biggr]\nonumber \\
&& \times \Biggl[1-\tilde F (Z,A ; \{Z_i, A_i\})^2\Biggr]^{-1}\, .
\end{eqnarray}
Using Eqs.~(\ref{eq:PL}), (\ref{eq:P}), and the expression for the pressure of an ideal electron Fermi gas (see, e.g., Chapter 2 in Ref.~\cite{hae07}), the threshold pressure at the onset of the phase transition is given by 
\begin{eqnarray}
\label{eq:exact-threshold-pressure}
P_{\rm thres}&=&\frac{m_e c^2}{8 \pi^2 \lambda_e^3} \biggl[x_r\left(\frac{2}{3}x_r^2-1\right)\sqrt{1+x_r^2}+\ln(x_r+\sqrt{1+x_r^2})\biggr]\nonumber\\
&&+\frac{C_{\rm bcc} \alpha}{3 (3\pi^2)^{4/3}} x_r^4 \frac{m_e c^2}{\lambda_e^3}Z^{2/3} 
\, .
\end{eqnarray}
The maximum mean nucleon number density $\bar n^{\rm max}$ up to which the pure solid of nuclei $(A,Z)$ is present is given by 
\begin{eqnarray}\label{eq:exact-threshold-density}
\bar n^{\rm max} = \frac{A}{Z} n_e= \frac{A}{Z} \frac{x_r^3 }{3 \pi^2 \lambda_e^3}\, .
\end{eqnarray}
The minimum possible mean nucleon number density $\bar n_{\{i\}}^{\rm min}$ at which the multi-component solid appears is given by 
\begin{eqnarray}\label{eq:exact-threshold-density-min}
\bar n_{\{i\}}^{\rm min} = \frac{\bar A}{\bar Z} (n_e+\delta n_e)  = \frac{\bar A}{\bar Z}\frac{Z}{A}\bar n^{\rm max} \Biggl[1+\frac{C_{\rm bcc} \alpha}{(3 \pi^2)^{1/3}} \biggl(Z^{2/3}-\tilde{f}(\{Z_i\})\biggr)\frac{\sqrt{1+x_r^2}}{x_r}\Biggr]\, . 
\end{eqnarray}
The transition is thus accompanied by a density discontinuity given by 
\begin{eqnarray}
\frac{\bar n_{\{i\}}^{\rm min}- \bar n^{\rm max}}{\bar n^{\rm max}} = \frac{\bar A}{\bar Z}\frac{Z}{A} \Biggl[1+\frac{C_{\rm bcc} \alpha}{(3 \pi^2)^{1/3}} \biggl(Z^{2/3}-\tilde{f}(\{Z_i\})\biggr)\frac{\sqrt{1+x_r^2}}{x_r}\Biggr]-1\, . 
\end{eqnarray}
According to Le Chatelier's principle, mechanical stability requires $\bar n_{\{i\}}^{\rm min}\geq \bar n^{\rm max}$. Since this constraint must be fulfilled irrespective of 
the small lattice correction, we thus obtain 
\begin{eqnarray}
{\bar A} Z - {\bar Z} A \geq 0 \, . 
\end{eqnarray}
In other words, the multi-component phase must be more neutron rich than the pure solid phase. 

In the regime of ultrarelativistic electrons such that $\gamma_e^{\rm thres} \gg 1$, Eq.~(\ref{eq:exact-xr}) reduces to 
\begin{eqnarray}
 x_r \approx \gamma_e^{\rm thres} \Biggl[1+\tilde F (Z,A ; \{Z_i, A_i\})\Biggr]^{-1}\, .
\end{eqnarray}
The threshold pressure~(\ref{eq:exact-threshold-pressure}), and the densities~(\ref{eq:exact-threshold-density}) and 
(\ref{eq:exact-threshold-density-min}) of the solid phases become respectively 
\begin{eqnarray}\label{eq:threshold-pressure}
P_{\rm thres} \approx &&\frac{(\mu_e^{\rm thres})^4}{12 \pi^2 (\hbar c)^3}\Biggl[1+\tilde F(Z,A;\{Z_i,A_i\})\Biggr]^{-4} \left(1+\frac{4 C_{\rm bcc} \alpha}{(81\pi^2)^{1/3}}Z^{2/3}\right)\, ,
\end{eqnarray}
\begin{eqnarray}\label{eq:threshold-density}
\bar n^{\rm max}  &\approx& \frac{A}{Z} \frac{(\mu_e^{\rm thres})^3}{3 \pi^2 (\hbar c)^3}\Biggl[1+
\tilde F(Z,A;\{Z_i,A_i\})\Biggr]^{-3}\, ,
\end{eqnarray}
\begin{eqnarray}\label{eq:threshold-density-min}
\bar n_{\{i\}}^{\rm min} \approx \frac{\bar A}{\bar Z}\frac{Z}{A}\bar n^{\rm max} \Biggl[1+\frac{C_{\rm bcc} \alpha}{(3 \pi^2)^{1/3}} \biggl(Z^{2/3}-\tilde{f}(\{Z_i\})\biggr)\Biggr]\, . 
\end{eqnarray}
The density discontinuity is thus approximately given by 
\begin{eqnarray}
\frac{\bar n_{\{i\}}^{\rm min}- \bar n^{\rm max}}{\bar n^{\rm max}} \approx  \frac{\bar A}{\bar Z}\frac{Z}{A} \Biggl[1+\frac{C_{\rm bcc} \alpha}{(3 \pi^2)^{1/3}} \biggl(Z^{2/3}-\tilde{f}(\{Z_i\})\biggr)\Biggr] -1\, . 
\end{eqnarray}

The formulas presented in this section remain valid at finite temperatures $T$ such that (i) electrons remain highly degenerate,
and (ii) matter in both phases is crystallized. The first condition requires $T\ll T_{{\rm F}e}$, where $T_{{\rm F}e}$ is the electron 
Fermi temperature defined by 
\begin{equation}\label{TFe}
T_{\text{F}e}=\frac{\mu_e-m_e c^2}{k_\text{B}}\, ,
\end{equation}
and $k_\text{B}$ is the Boltzmann's constant. As for the second condition, we must have $T < T_m^0$ and $T< T_m$, where $T_m^0$ and $T_m$ denote the 
crystallization temperatures of the pure and multicomponents phases, as defined by (see, e.g., Ref.~\cite{hae07})
\begin{equation}\label{eq:Tm}
T_m=\frac{e^2}{a_e k_\text{B} \Gamma_m}\overline{Z^{5/3}} \, ,
\end{equation}
and similarly for $T_m^0$, 
where $a_e$ is the electron-sphere radius, $\Gamma_m$ is 
the Coulomb coupling parameter at melting, and $\overline{Z^{5/3}}=\sum_i \xi_i Z_i^{5/3}$. 
Neglecting the lattice correction in Eq.~(\ref{eq:threshold-condition}) such that $\mu_e \approx \mu_e^{\rm thres}$, and 
assuming electrons are ultrarelativistic, the electron Fermi temperature and the crystallization temperature 
can be approximately expressed as 
\begin{equation}
T_{{\rm F}e}\approx 5.93\times 10^9  \frac{\mu_e^{\rm thres}}{m_e c^2}~\text{K}\, ,
\end{equation}
\begin{equation}
T_m \approx 1.29 \times 10^5 \frac{\mu_e^{\rm thres}}{m_e c^2} \overline{Z^{5/3}}~\text{K}\, ,
\end{equation}
where we have adopted the value $\Gamma_m=175$ for the Coulomb coupling parameter at melting~\cite{hae07}. 
Since typically $\overline{Z^{5/3}}$ and $Z^{3/5}$ are of order $\sim 300-500$ 
(see, e.g. Tables~\ref{tab:hfb24-pure} and \ref{tab:hfb24-compound}), 
we thus have $T_m^0 \sim T_m \ll T_{{\rm F}e}$.

\subsection{Formation of a solid compound}
\label{sec:compound}

A multi-component solid may consist of (i) spatially separated pure bcc phases, or (ii) a compound. The latter may not necessarily form for arbitrary composition. 
Moreover, a compound may be ordered or disordered depending on the charges $\{Z_i\}$ (see, e.g. Ref.~\cite{iga01}). 

A compound made of nuclei $(A_i,Z_i)$ is stable against the separation into pure bcc phases if the Gibbs free energy per nucleon 
of the compound is lower than that of the coexisting phases at the \emph{same} pressure $P$ and at the \emph{same} composition. 
Neglecting surface effects, the Gibbs free energy per nucleon of coexisting phases can be written in the form (\ref{eq:gibbs}) after substituting the structure 
function $f(\{Z_i\})$ by (see, e.g., Section 2.4.7 in Ref.~\cite{hae07})
\begin{equation}\label{eq:fmix}
 \tilde{f}_{\rm mix}(\{Z_i\}) = f_{\rm mix}(\{Z_i\})= \frac{ \overline{Z^{5/3}}}{\bar Z}\, .
\end{equation}
Expanding $g$ to first order in $\alpha$, the stability of a multinary compound against phase separation can be expressed as 
\begin{equation}\label{eq:separation}
 \mathcal{R}\equiv \frac{\tilde{f}(\{Z_i\})}{f_{\rm mix}(\{Z_i\})} > 1\, ,
\end{equation}
irrespective of the stellar conditions. In other words, if a compound with a \emph{fixed} composition is found to be stable at some pressure $P$, 
it will remain so at any other pressure and independently of the degree of relativity of the electron gas, as recently noticed in Ref.~\cite{eng16}.
The condition~(\ref{eq:separation}) generalizes that originally obtained by Dyson~\cite{dyson71} in the case of a binary compound 
under the approximation $P\approx P_e$, see his Eq.(1.17). In particular, our present derivation shows that Eq.~(\ref{eq:separation}) still 
remains valid if the lattice contribution to the pressure is taken into account. Let us emphasize that this inequality only pertains to the stability 
of a multinary compound against the formation of pure coexisting bcc phases, and does not not preclude the occurrence of instabilities due to weak 
and strong nuclear processes, as we shall show in Section~\ref{sec:composition}. 

In full thermodynamic equilibrium, as generally assumed in the crust of nonaccreting neutron stars~\cite{hw58,htww65}, the formation of multinary compounds  
made of a large variety of different nuclear species seems unlikely (see, e.g., Ref.~\cite{gul15}). In what follows, we shall thus focus on binary compounds 
since they could be found at the interface between adjacent layers of neutron-star crusts~\cite{jog82}. We shall also briefly discuss the existence of 
ternary compounds in the crust. The possible charge ratios of these compounds are expected to remain close to $Z_2/Z_1\sim 1$ according to crustal compositions predicted 
by recent models (see, e.g., Refs.~\cite{roca2008,pearson2011,hemp2013,chamel2015c,utama2016}; see also Table~\ref{tab:hfb24-pure}). For such charge ratios, 
disordered compounds are unstable~\cite{oga93,iga01}, and therefore will not be further considered.

\section{Transitions between one and two-component solid phases} 
\label{sec:transition2}

\subsection{Transitions between two pure solid phases}
\label{sec:trans-pure}

\begin{figure}
 \centering
 \includegraphics[scale=.85]{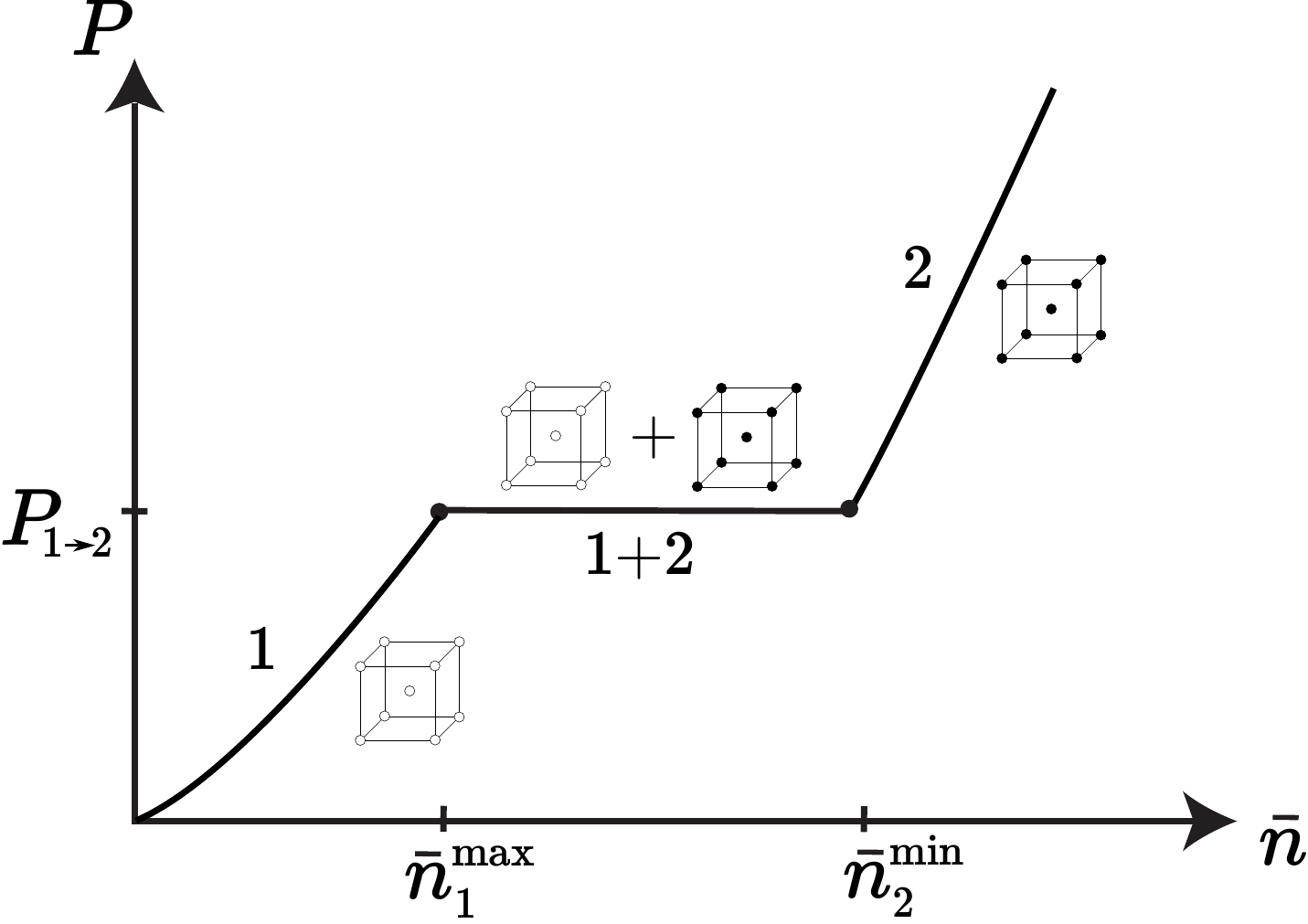}
 \caption{Schematic representation of the pressure $P$ versus mean baryon number density $\bar n$ for a transition between two pure body-centered cubic lattice solid 
 phases of nuclei $(A_1,Z_1)$ and $(A_2,Z_2)$. The two phases coexist at pressure $P_{1\rightarrow2}$ irrespective of their proportion. }
 \label{fig:transitions-pure}
\end{figure}

Let us consider as a limiting case of the general situation considered in the previous section, the transition between two pure solid phases, 
made of nuclei $(A_1,Z_1)$ and $(A_2,Z_2)$ respectively (arranged in a bcc lattice, as discussed earlier). The variation of the pressure with 
respect to the density is schematically illustrated in Fig.~\ref{fig:transitions-pure}. 

The threshold conditions can be readily obtained using the formulas given in Section~\ref{sec:threshold-conditions}. 
The relativity parameter is thus given by 
\begin{eqnarray}
 x_r&=&\gamma_e^{1\rightarrow 2} \Biggl[\sqrt{1-\left(1-\tilde F_0(Z_1, A_1;Z_2,A_2)^2\right)/(\gamma_e^{1\rightarrow2})^{2}}-\tilde F_0(Z_1, A_1; Z_2,A_2)\Biggr]\nonumber \\
&& \times \Biggl[1-\tilde F_0 (Z_1, A_1; Z_2,A_2)^2\Biggr]^{-1}\, .
\end{eqnarray}
with 
\begin{equation}\label{eq:gammae-1>2}
\gamma_e^{1\rightarrow2}\equiv \frac{\mu_e^{1\rightarrow2}}{m_e c^2}\, , 
\end{equation}
\begin{equation}\label{eq:mue-pure}
\mu_e^{1\rightarrow2}\equiv \biggl[\frac{M^\prime(A_2,Z_2)c^2}{A_2}-\frac{M^\prime(A_1,Z_1)c^2}{A_1}\biggr]\left(\frac{Z_1}{A_1}-\frac{Z_2}{A_2}\right)^{-1} +  m_e c^2\, ,
\end{equation}
\begin{eqnarray}
\tilde F_0(Z_1,A_1;Z_2,A_2) &\equiv& \tilde F(Z_1,A_1;Z_2,A_2,Z_2,A_2) \nonumber \\
 &=& \frac{C_{\rm bcc}}{(3\pi^2)^{1/3}}\alpha\left(\frac{4}{3}\frac{Z_1^{5/3}}{A_1} - \frac{1}{3}\frac{Z_1^{2/3} Z_2}{A_2} -\frac{Z_2^{5/3}}{A_2}\right)\left(\frac{Z_1}{A_1}-\frac{Z_2}{A_2}\right)^{-1}\, .
\end{eqnarray}
The threshold pressure $P_{1\rightarrow2}$ is given by 
\begin{eqnarray}
\label{eq:exact-threshold-pressure-1>2}
P_{1\rightarrow2}&=&\frac{m_e c^2}{8 \pi^2 \lambda_e^3} \biggl[x_r\left(\frac{2}{3}x_r^2-1\right)\sqrt{1+x_r^2}+\ln(x_r+\sqrt{1+x_r^2})\biggr]\nonumber\\
&&+\frac{C_{\rm bcc} \alpha}{3 (3\pi^2)^{4/3}} x_r^4 \frac{m_e c^2}{\lambda_e^3}Z_1^{2/3} 
\, .
\end{eqnarray}
The highest possible density at which nuclei $(A_1,Z_1)$ are stable is given by 
\begin{eqnarray}
\bar n_1^{\rm max} = \frac{A_1}{Z_1} \frac{x_r^3 }{3 \pi^2 \lambda_e^3}\, ,
\end{eqnarray}
while the lowest possible density at which nuclei $(A_2,Z_2)$ can be found is given by 
\begin{eqnarray}
\bar n_{2}^{\rm min} = \frac{A_2}{Z_2}\frac{Z_1}{A_1}\bar n_1^{\rm max} \Biggl[1+\frac{C_{\rm bcc} \alpha}{(3 \pi^2)^{1/3}} \biggl(Z_1^{2/3}-Z_2^{2/3}\biggr)
\frac{\sqrt{1+x_r^2}}{x_r}\Biggr]\, . 
\end{eqnarray}
The transition is thus accompanied by a density discontinuity given by 
\begin{eqnarray}\label{eq:exact-density-discontinuity-1>2}
\frac{\bar n_{2}^{\rm min}- \bar n_1^{\rm max}}{\bar n_1^{\rm max}} = \frac{A_2}{Z_2}\frac{Z_1}{A_1} 
\Biggl[1+\frac{C_{\rm bcc} \alpha}{(3 \pi^2)^{1/3}} \biggl(Z_1^{2/3}-Z_2^{2/3}\biggr)\frac{\sqrt{1+x_r^2}}{x_r}\Biggr]-1\, . 
\end{eqnarray}
As discussed at the end of Section~\ref{sec:threshold-conditions}, mechanical stability requires  
\begin{eqnarray}\label{eq:mech12}
A_2 Z_1 - Z_2 A_1 \geq 0 \, . 
\end{eqnarray}

In the ultrarelativistic regime $\gamma_e^{1\rightarrow2}\gg1$, the threshold pressure and the densities of the two solid phases can be approximately expressed as 
\begin{eqnarray}\label{eq:threshold-pressure-1>2}
P_{1\rightarrow2} \approx &&\frac{(\mu_e^{1\rightarrow2})^4}{12 \pi^2 (\hbar c)^3}\Biggl[1+\tilde F_0(Z_1,A_1;Z_2,A_2)\Biggr]^{-4} 
\left(1+\frac{4 C_{\rm bcc} \alpha}{(81\pi^2)^{1/3}}Z_1^{2/3}\right)\, ,
\end{eqnarray}
\begin{eqnarray}
\bar n_1^{\rm max}  &\approx& \frac{A_1}{Z_1} \frac{(\mu_e^{1\rightarrow2})^3}{3 \pi^2 (\hbar c)^3}\Biggl[1+
\tilde F_0(Z_1,A_1;Z_2,A_2)\Biggr]^{-3}\, ,
\end{eqnarray}
\begin{eqnarray}
\bar n_{2}^{\rm min} \approx \frac{A_2}{Z_2}\frac{Z_1}{A_1}\bar n_1^{\rm max} \Biggl[1+\frac{C_{\rm bcc} \alpha}{(3 \pi^2)^{1/3}} \biggl(Z_1^{2/3}-Z_2^{2/3}\biggr)\Biggr]\, . 
\end{eqnarray}
The density discontinuity is thus approximately given by 
\begin{eqnarray}
\frac{\bar n_{2}^{\rm min}- \bar n_1^{\rm max}}{\bar n_1^{\rm max}} \approx  \frac{A_2}{Z_2}\frac{Z_1}{A_2} \Biggl[1+\frac{C_{\rm bcc} \alpha}{(3 \pi^2)^{1/3}} 
\biggl(Z_1^{2/3}-Z_2^{2/3}\biggr)\Biggr] -1\, .
\end{eqnarray}

The apparent dissymmetry of Eq.~(\ref{eq:exact-threshold-pressure-1>2}) hence also of Eq.~(\ref{eq:threshold-pressure-1>2}) 
under the interchange $1\leftrightarrow 2$ arises from the expansion of the Gibbs free 
energy per nucleon to first order in $\alpha$. The symmetry can be restored by expressing the pressure at first order in $\alpha$:
\begin{eqnarray}
P_{1\rightarrow2}&\approx &\frac{m_e c^2}{8 \pi^2 \lambda_e^3} \biggl[x_{r0}\left(\frac{2}{3}x_{r0}^2-1\right)\sqrt{1+x_{r0}^2}+\ln(x_{r0}+\sqrt{1+x_{r0}^2})\biggr]\nonumber\\
&&+\frac{C_{\rm bcc} \alpha}{(3\pi^2)^{4/3}} \frac{m_e c^2}{\lambda_e^3} x_{r0}^4 \frac{A_1 Z_2^{5/3}-A_2 Z_1^{5/3}}{A_2 Z_1-A_1 Z_2} 
\, ,
\end{eqnarray}
with 
\begin{equation}
x_{r0}=\sqrt{(\gamma_e^{1\rightarrow2})^2-1}\, .
\end{equation}
To this order, the pressure~(\ref{eq:exact-threshold-pressure-1>2}) can thus be equivalently written as 
\begin{eqnarray}
\label{eq:exact-threshold-pressure-1>2bis}
P_{1\rightarrow2}&=&\frac{m_e c^2}{8 \pi^2 \lambda_e^3} \biggl[x_r\left(\frac{2}{3}x_r^2-1\right)\sqrt{1+x_r^2}+\ln(x_r+\sqrt{1+x_r^2})\biggr]\nonumber\\
&&+\frac{C_{\rm bcc} \alpha}{3 (3\pi^2)^{4/3}} x_r^4 \frac{m_e c^2}{\lambda_e^3}Z_2^{2/3} 
\, ,
\end{eqnarray}
with 
\begin{eqnarray}
 x_r&=&\gamma_e^{1\rightarrow 2} \Biggl[\sqrt{1-\left(1-\tilde F_0(Z_2, A_2;Z_1,A_1)^2\right)/(\gamma_e^{1\rightarrow2})^{2}}-\tilde F_0(Z_2, A_2; Z_1,A_1)\Biggr]\nonumber \\
&& \times \Biggl[1-\tilde F_0 (Z_2, A_2; Z_1,A_1)^2\Biggr]^{-1}\, .
\end{eqnarray}
In the limit of ultrarelativistic electrons, we find 
\begin{eqnarray}\label{eq:P12-pure}
P_{1\rightarrow2} \approx \frac{(\mu_e^{1\rightarrow2})^4}{12 \pi^2 (\hbar c)^3}\Biggl(1-\frac{4 C_{\rm bcc}\alpha}{(3\pi^2)^{1/3}}
\frac{A_1 Z_2^{5/3} - A_2 Z_1^{5/3}}{A_1 Z_2 - A_2 Z_1}\Biggr)\, .
\end{eqnarray}

\subsection{Transitions from a pure solid phase of nuclei $(A_1,Z_1)$ to a two-component solid phase of nuclei $(A_1,Z_1)$ and $(A_2,Z_2)$}
\label{sec:trans-mix}

Using the formulas given in Section~\ref{sec:threshold-conditions}, the threshold pressure at the onset of the transition is given by 
\begin{eqnarray}
\label{eq:exact-threshold-pressure-1>1+2}
P_{1\rightarrow1+2}&=&\frac{m_e c^2}{8 \pi^2 \lambda_e^3} \biggl[x_r\left(\frac{2}{3}x_r^2-1\right)\sqrt{1+x_r^2}+\ln(x_r+\sqrt{1+x_r^2})\biggr]\nonumber\\
&&+\frac{C_{\rm bcc} \alpha}{3 (3\pi^2)^{4/3}} x_r^4 \frac{m_e c^2}{\lambda_e^3}Z_1^{2/3} 
\, ,
\end{eqnarray}
where 
\begin{eqnarray}
 x_r&=&\gamma_e^{1\rightarrow2} \Biggl[\sqrt{1-\left(1-\tilde F(Z_1,A_1 ; Z_1, A_1, Z_2,A_2)^2\right)/(\gamma_e^{1\rightarrow2})^{2}}-\tilde F(Z_1,A_1 ; Z_1, A_1, Z_2,A_2)\Biggr]
 \nonumber \\ && \times \Biggl[1-\tilde F (Z_1,A_1 ; Z_1, A_1, Z_2,A_2)^2\Biggr]^{-1}\, .
\end{eqnarray}
The threshold electron Fermi energy~(\ref{eq:muethres}) turns out be the same as that given by Eq.~(\ref{eq:mue-pure}) 
for the transition between pure bcc solid phases. 
However, $P_{1\rightarrow1+2}$ is generally not equal to $P_{1\rightarrow2}$ due to the different lattice contributions. 
The highest density at which nuclei $(A_1,Z_1)$ can be possibly present is given by 
\begin{eqnarray}
\bar n_1^{\rm max} = \frac{A_1}{Z_1} \frac{x_r^3 }{3 \pi^2 \lambda_e^3}\, .
\end{eqnarray}
The lowest density at which a two-component solid phase of nuclei $(A_1,Z_1)$ and $(A_2,Z_2)$ possibly appears is given by 
\begin{eqnarray}
\bar n_{1+2}^{\rm min} = \frac{\bar A}{\bar Z}\frac{Z_1}{A_1}\bar n_1^{\rm max} \Biggl[1+\frac{C_{\rm bcc} \alpha}{(3 \pi^2)^{1/3}} 
\biggl(Z_1^{2/3}-\tilde{f}(Z_1,Z_2)\biggr)\frac{\sqrt{1+x_r^2}}{x_r}\Biggr]\, . 
\end{eqnarray}
% The expressions for the other solid-solid transitions discussed in the articles can be easily obtained. 
The transition is thus accompanied by a density discontinuity given by 
\begin{eqnarray}\label{eq:exact-density-discontinuity-1>1+2}
\frac{\bar n_{1+2}^{\rm min}- \bar n_1^{\rm max}}{\bar n_1^{\rm max}} = \frac{\bar A}{\bar Z}\frac{Z_1}{A_1} 
\Biggl[1+\frac{C_{\rm bcc} \alpha}{(3 \pi^2)^{1/3}} \biggl(Z_1^{2/3}-\tilde{f}(Z_1,Z_2)\biggr)\frac{\sqrt{1+x_r^2}}{x_r}\Biggr]-1\, . 
\end{eqnarray}
As discussed at the end of Section~\ref{sec:threshold-conditions}, we must have 
\begin{eqnarray}\label{eq:lechatelier1}
\bar A Z_1 - \bar Z A_1 \geq 0 \, ,
\end{eqnarray}
to ensure mechanical stability. 

Assuming electrons are ultrarelativistic, i.e. $\gamma_e^{1\rightarrow2}\gg1$, the threshold pressure and the densities of the two solid phases 
are approximately given by 
\begin{eqnarray}\label{eq:P1>1+2}
P_{1\rightarrow1+2} \approx \frac{(\mu_e^{1\rightarrow2})^4}{12 \pi^2 (\hbar c)^3}\Biggl[1+
\tilde F(Z_1,A_1;Z_1,A_1,Z_2,A_2)\Biggr]^{-4} \left(1+\frac{4 C_{\rm bcc} \alpha}{(81\pi^2)^{1/3}}Z_1^{2/3}\right)\, ,
\end{eqnarray}
\begin{eqnarray}
\bar n_1^{\rm max} \approx \frac{A_1}{Z_1} \frac{(\mu_e^{1\rightarrow2})^3}{3 \pi^2 (\hbar c)^3}\Biggl[1+
\tilde F(Z_1,A_1;Z_1,A_1,Z_2,A_2)\Biggr]^{-3}\, ,
\end{eqnarray}
\begin{eqnarray}
\bar n_{1+2}^{\rm min} \approx \frac{\bar A}{\bar Z}\frac{Z_1}{A_1} \bar n_1^{\rm max}
\biggl[ 1 + \frac{C_{\rm bcc}\alpha}{(3 \pi^2)^{1/3}}  \biggl(Z_1^{2/3}-\tilde{f}(Z_1,Z_2)\biggr)\biggr]\, .
\end{eqnarray}
The density discontinuity is thus approximately given by 
\begin{eqnarray}\label{eq:dn12-mix}
\frac{\bar n_{1+2}^{\rm min}- \bar n_1^{\rm max}}{\bar n_1^{\rm max}} \approx \frac{\bar A}{\bar Z}\frac{Z_1}{A_1} 
\biggl[ 1 + \frac{C_{\rm bcc}\alpha}{(3 \pi^2)^{1/3}}  \biggl(Z_1^{2/3}-\tilde{f}(Z_1,Z_2)\biggr)\biggr]-1\, .
\end{eqnarray}

\subsection{Coexistence of two pure solid phases}
\label{sec:coexistence}

The onset of the transition from a pure bcc phase of nuclei $(A_1,Z_1)$ to a coexistence of 
bcc phases of nuclei $(A_1,Z_1)$ and $(A_2,Z_2)$ is found to be determined by the \emph{same} stability condition as that for the transition between 
two pure bcc phases of nuclei $(A_1,Z_1)$ and $(A_2,Z_2)$ respectively.
In particular, the threshold electron Fermi energy and the pressure are still given by Eqs.~(\ref{eq:mue-pure}) and (\ref{eq:exact-threshold-pressure-1>2}) 
respectively. However, the density now varies \emph{continuously} at the transition, as can be seen from Eq.~(\ref{eq:exact-density-discontinuity-1>1+2}) with $\xi=1$, 
recalling that $\tilde{f}_{\rm mix}(Z_1,Z_2)= \overline{Z^{5/3}}/\bar Z$ in this case (see Section~\ref{sec:compound}). 
As $\xi$ decreases from $\xi=1$ to $\xi=0$, the mean density $\bar n$ increases from $\bar n_1^{\rm max}$ (pure bcc phase of nuclei $(A_1,Z_1)$) to 
$\bar n_2^{\rm min}$ (pure bcc phase of nuclei $(A_2,Z_2)$),  whereas the pressure remains unchanged $P=P_{1\rightarrow2}$, as illustrated in 
Fig.~\ref{fig:transitions-pure}. Since inside a self-gravitating body in hydrostatic equilibrium the pressure must increase monotonically with depth 
(see, e.g., Ref.~\cite{hae07}), pure solid phases cannot coexist in any region of the crust of a neutron star.

\subsection{Stability of a binary compound against phase separation}
\label{sec:compound2}

As discussed in Section~\ref{sec:compound}, the stability of a binary compound made of nuclei $(A_1,Z_1)$ and $(A_2,Z_2)$ against phase separation is determined by the condition~(\ref{eq:separation}), which reads 
\begin{equation}\label{eq:separation2}
 \mathcal{R}\equiv \frac{\tilde{f}(Z_1,Z_2)}{f_{\rm mix}(Z_1,Z_2)} > 1\, .
\end{equation}
If fulfilled, the inequality~(\ref{eq:separation2}) implies that a mixture of two pure bcc phases of nuclei 
$(A_1,Z_1)$ and $(A_2,Z_2)$ 
at pressure $P_{1\rightarrow2}$ is unstable against the formation of a binary compound. As a consequence, the threshold pressure $P_{1\rightarrow1+2}$ for the 
appearance of the compound must be lower than $P_{1\rightarrow2}$. To show this, let us first remark that 
\begin{eqnarray}
\tilde F(Z_1,A_1;Z_1,A_1,Z_2,A_2)-\tilde F_0(Z_1,A_1;Z_2,A_2)=A_1 \frac{\overline{Z^{5/3}}-\tilde{f}(Z_1,Z_2) \bar Z}{Z_1 A_2-A_1 Z_2} \frac{C_{\rm bcc}\alpha}{(3 \pi^2)^{1/3}} \, . 
\end{eqnarray}
Using Eqs.~(\ref{eq:fmix}), (\ref{eq:mech12}), and (\ref{eq:separation2}), and recalling that $C_{\rm bcc}<0$, we thus have
\begin{eqnarray}
\tilde F(Z_1,A_1;Z_1,A_1,Z_2,A_2) > \tilde F_0(Z_1,A_1;Z_2,A_2)\, . 
\end{eqnarray}
The inequality $P_{1\rightarrow1+2}<P_{1\rightarrow2}$ follows by comparing Eqs.~(\ref{eq:exact-threshold-pressure-1>2}) and (\ref{eq:exact-threshold-pressure-1>1+2}). 
Likewise, the highest possible density $\bar n_1^{\rm max}$ at which the pure bcc solid phase of nuclei $(A_1,Z_1)$ can possibly exist is lower than that obtained 
for the transition between the two pure solid phases if Eq.~(\ref{eq:separation2}) holds. 

With further compression, the compound will be unstable against the transition to a pure bcc phase of nuclei $(A_2,Z_2)$. The pressure $P_{1+2\rightarrow2}$ 
at which this transition occurs can be obtained using the formulas given in Section~\ref{sec:threshold-conditions}:  
\begin{eqnarray}
\label{eq:exact-threshold-pressure-2>1+2}
P_{1+2\rightarrow2}&=&\frac{m_e c^2}{8 \pi^2 \lambda_e^3} \biggl[x_r\left(\frac{2}{3}x_r^2-1\right)\sqrt{1+x_r^2}+\ln(x_r+\sqrt{1+x_r^2})\biggr]\nonumber\\
&&+\frac{C_{\rm bcc} \alpha}{3 (3\pi^2)^{4/3}} x_r^4 \frac{m_e c^2}{\lambda_e^3}Z_2^{2/3} 
\, ,
\end{eqnarray}
\begin{eqnarray}
 x_r&=&\gamma_e^{1\rightarrow2} \Biggl[\sqrt{1-\left(1-\tilde F(Z_2,A_2 ; Z_1, A_1, Z_2,A_2)^2\right)/(\gamma_e^{1\rightarrow2})^{2}}-\tilde F(Z_2,A_2 ; Z_1, A_1, Z_2,A_2)
 \Biggr]
 \nonumber \\ && \times \Biggl[1-\tilde F (Z_2,A_2 ; Z_1, A_1, Z_2,A_2)^2\Biggr]^{-1}\, ,
\end{eqnarray}
with the same threshold electron Fermi energy as that given by Eq.~(\ref{eq:mue-pure}) for the transition between pure bcc solid phases.
The highest density at which the compound is possibly present is given by 
\begin{eqnarray}
\bar n_{1+2}^{\rm max} = \frac{\bar A}{\bar Z}\frac{Z_2}{A_2}\bar n_2^{\rm max} \Biggl[1+\frac{C_{\rm bcc} \alpha}{(3 \pi^2)^{1/3}} 
\biggl(Z_2^{2/3}-\tilde{f}(Z_1,Z_2)\biggr)\frac{\sqrt{1+x_r^2}}{x_r}\Biggr]\, ,
\end{eqnarray}
where 
\begin{eqnarray}
\bar n_2^{\rm min} = \frac{A_2}{Z_2} \frac{x_r^3 }{3 \pi^2 \lambda_e^3}
\end{eqnarray}
is the lowest density at which the pure bcc solid phase of nuclei $(A_2,Z_2)$ can appear. 
The transition between these two phases is accompanied by a density discontinuity given by 
\begin{eqnarray}\label{eq:exact-density-discontinuity-2>1+2}
\frac{\bar n_{1+2}^{\rm max}- \bar n_2^{\rm min}}{\bar n_2^{\rm min}} = \frac{\bar A}{\bar Z}\frac{Z_2}{A_2} 
\Biggl[1+\frac{C_{\rm bcc} \alpha}{(3 \pi^2)^{1/3}} \biggl(Z_2^{2/3}-\tilde{f}(Z_1,Z_2)\biggr)\frac{\sqrt{1+x_r^2}}{x_r}\Biggr]-1\, . 
\end{eqnarray}
As discussed at the end of Section~\ref{sec:threshold-conditions}, we must have 
\begin{eqnarray}\label{eq:lechatelier2}
\bar A Z_2 - \bar Z A_2 \leq 0 \, ,
\end{eqnarray}
to ensure mechanical stability. Collecting Eqs.~(\ref{eq:lechatelier1}) and (\ref{eq:lechatelier2}) thus shows that the composition of the compound is not arbitrary, 
but must satisfy the following constraint (matter neutronization): 
\begin{eqnarray}\label{eq:lechatelier1+2}
\frac{A_1}{Z_1}\leq\frac{\bar A}{\bar Z} \leq \frac{A_2}{Z_2} \, .
\end{eqnarray}

In the limit of ultrarelativistic electrons, $\gamma_e^{1\rightarrow2}\gg1$, the threshold pressure and the associated densities are 
approximately given by 
\begin{eqnarray}\label{eq:P2>1+2}
P_{1+2\rightarrow2} \approx \frac{(\mu_e^{1\rightarrow2})^4}{12 \pi^2 (\hbar c)^3}\Biggl[1+
\tilde F(Z_2,A_2;Z_1,A_1,Z_2,A_2)\Biggr]^{-4}\left(1+\frac{4 C_{\rm bcc} \alpha}{(81\pi^2)^{1/3}}Z_2^{2/3}\right)\, ,
\end{eqnarray}
\begin{eqnarray}\label{eq:n1+2max}
\bar n_{1+2}^{\rm max} \approx \frac{\bar A}{\bar Z}\frac{Z_2}{A_2} \bar n_2^{\rm min}\biggl[ 1 + \frac{C_{\rm bcc}\alpha}{(3 \pi^2)^{1/3}}  \biggl(Z_2^{2/3}-\tilde{f}(Z_1,Z_2)\biggr)\biggr]\, ,
\end{eqnarray}
\begin{eqnarray}
\bar n_2^{\rm min} \approx \frac{A_2}{Z_2} \frac{(\mu_e^{1\rightarrow2})^3}{3 \pi^2 (\hbar c)^3}\Biggl[1+
\tilde F(Z_2,A_2;Z_1,A_1,Z_2,A_2)\Biggr]^{-3}\, .
\end{eqnarray}
The density discontinuity is thus approximately given by 
\begin{eqnarray}\label{eq:dn21-mix}
\frac{\bar n_{1+2}^{\rm max}- \bar n_2^{\rm min}}{\bar n_2^{\rm min}} \approx \frac{\bar A}{\bar Z}\frac{Z_2}{A_2} 
\biggl[ 1 + \frac{C_{\rm bcc}\alpha}{(3 \pi^2)^{1/3}}  \biggl(Z_2^{2/3}-\tilde{f}(Z_1,Z_2)\biggr)\biggr]-1\, .
\end{eqnarray}

If the binary compound is stable, i.e. Eq.~(\ref{eq:separation2}) is fulfilled, it can be shown that $P_{1+2\rightarrow2}>P_{1\rightarrow2}$ by comparing 
Eqs.~(\ref{eq:exact-threshold-pressure-1>2bis}) and (\ref{eq:exact-threshold-pressure-2>1+2}) using Eq.~(\ref{eq:mech12}) and the identity
\begin{eqnarray}
\tilde F(Z_2,A_2;Z_1,A_1,Z_2,A_2) - \tilde F_0(Z_2,A_2;Z_1,A_1)=A_2 \frac{\overline{Z^{5/3}}-\bar Z \tilde{f}(Z_1,Z_2)}{Z_2 A_1 - Z_1 A_2} \frac{C_{\rm bcc}\alpha}{(3\pi^2)^{1/3}} \, .
\end{eqnarray}
Similarly, it can be shown that $\bar n^{\rm max}_{1+2} > \bar n^{\rm min}_{1+2}$. Binary compounds can exist in the crust of a neutron star, but only over a 
very small range of pressures, which to lowest order in $\alpha$ is approximately given by 
\begin{equation}\label{eq:dP1+2}
%  \frac{P_{1+2\rightarrow2}-P_{1\rightarrow1+2}}{P_{1\rightarrow2}}\approx \frac{4 C_{\rm bcc}\alpha}{(3\pi^2)^{1/3}} \bar A \frac{\bar Z \tilde{f}(Z_1,Z_2)-\overline{Z^{5/3}}}{\xi(1-\xi)(A_1 Z_2 - A_2 Z_1)}\, .
  \frac{P_{1+2\rightarrow2}-P_{1\rightarrow1+2}}{P_{1\rightarrow2}}\approx \frac{4 C_{\rm bcc}\alpha}{(3\pi^2)^{1/3}}  \frac{\bar A \bar Z (A_2 Z_1 - A_1 Z_2)}{(Z_1 \bar A - \bar Z A_1)(Z_2 \bar A-\bar Z A_2)} \left(\tilde{f}(Z_1,Z_2)-\frac{\overline{Z^{5/3}}}{\bar Z}\right)\, , %CHECK
\end{equation}
where we have used Eqs.~(\ref{eq:P12-pure}), (\ref{eq:P1>1+2}), (\ref{eq:P2>1+2}). This equation also shows that compound made of different isotopes are unlikely to be 
present in the crust since $P_{1+2\rightarrow2} = P_{1\rightarrow1+2}$ if $Z_1=Z_2$. The range of densities for which the compound exists is approximately given to lowest order in 
$\alpha$ by 
\begin{equation}\label{eq:dn1+2}
 \frac{\bar n_{1+2}^{\rm max}-\bar n_{1+2}^{\rm min}}{\bar n_{2}^{\rm min} - \bar n_1^{\rm max}}\approx \frac{3 C_{\rm bcc}\alpha}{(3\pi^2)^{1/3}}
 \left(\tilde{f}(Z_1,Z_2)-\frac{\overline{Z^{5/3}}}{\bar Z}\right)\left(1-\frac{\bar Z A_1}{\bar A Z_1}\right)^{-1}\left(1-\frac{\bar Z A_2}{\bar A Z_2}\right)^{-1} \, .
\end{equation}

\begin{figure}
 \centering
 \includegraphics[scale=0.85]{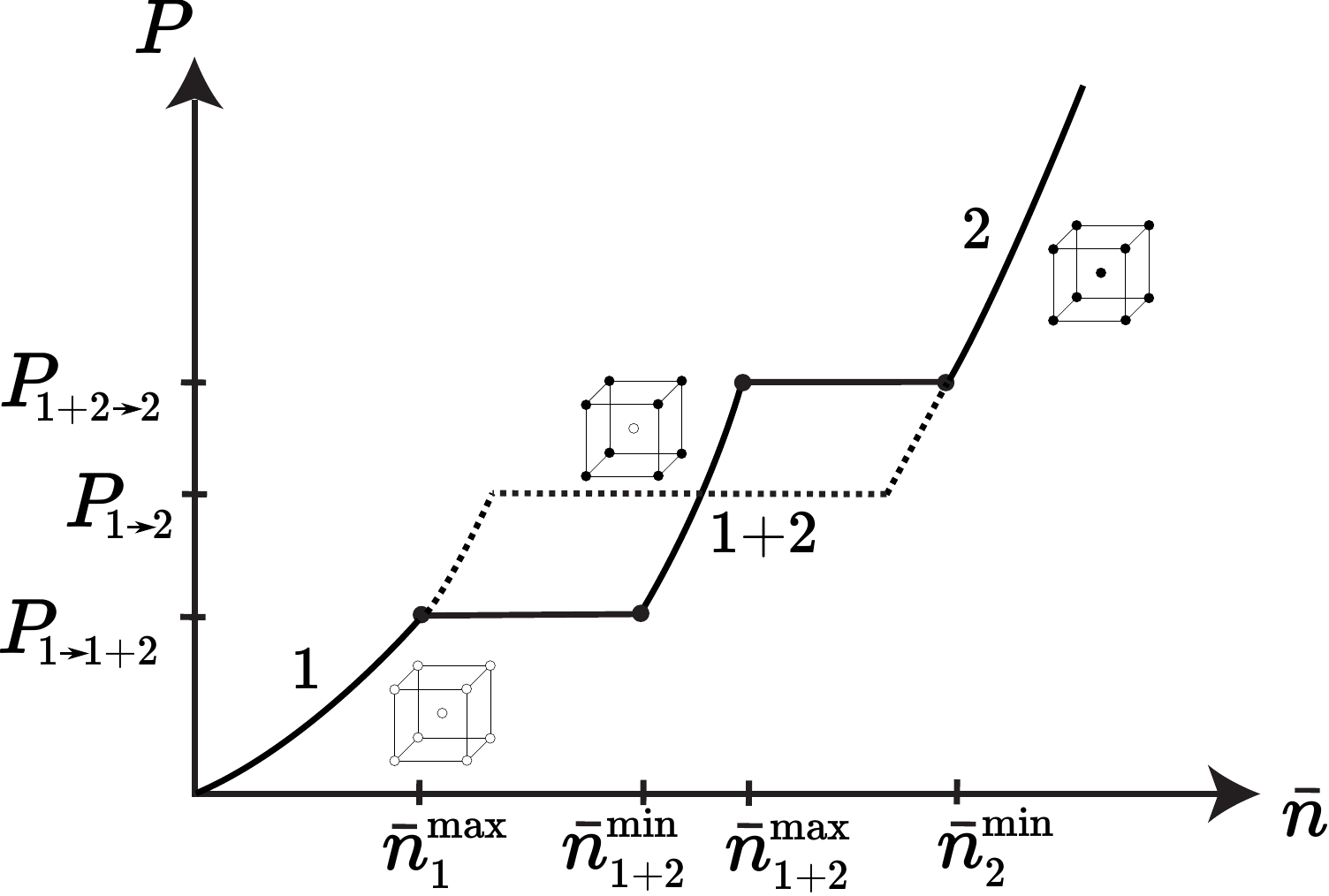}
 \caption{Schematic representation of the pressure $P$ versus mean baryon number density $\bar n$ for a transition between two pure body-centered cubic solid phases of nuclei 
 $(A_1,Z_1)$ and $(A_2,Z_2)$ accompanied by the formation of a binary compound (with cesium chloride structure in this case). For comparison, the transition leading 
 to the coexistence of pure phases is 
 indicated by the dotted line. The figure is not to scale. In reality, the range of pressures for which the compound is present (from $P_{1\rightarrow1+2}$ to 
 $P_{1+2\rightarrow2}$) is very small, $(P_{1+2\rightarrow2} - P_{1\rightarrow1+2})/ P_{1\rightarrow2} \ll 1$, as shown in Eq.~(\ref{eq:dP1+2}). 
 }
 \label{fig:transitions}
\end{figure}

The variation of the pressure with respect to the density is schematically illustrated in Fig.~\ref{fig:transitions}. 
In principle, the pure bcc crystal of nuclei  $(A_1,Z_1)$  and the binary compound can coexist at pressure $P_{1\rightarrow1+2}$ (at densities between $\bar n_1^{\rm max}$ 
and $\bar n_{1+2}^{\rm min}$); similarly the pure bcc crystal of nuclei  $(A_2,Z_2)$  and the binary compound can coexist at pressure $P_{1+2\rightarrow2}$ (at densities 
between $\bar n_{1+2}^{\rm max}$ and $\bar n_{2}^{\rm min}$). However, as discussed in Section~\ref{sec:coexistence}, such coexistence of solid phases cannot occur in any 
region of neutron-star crusts. 

\section{Equilibrium composition of the outer crust of cold nonaccreting neutron stars}
\label{sec:composition}

\subsection{Stability of various cubic and noncubic binary compounds against phase separation}

\begin{figure*}
\begin{center}
\includegraphics[scale=0.35]{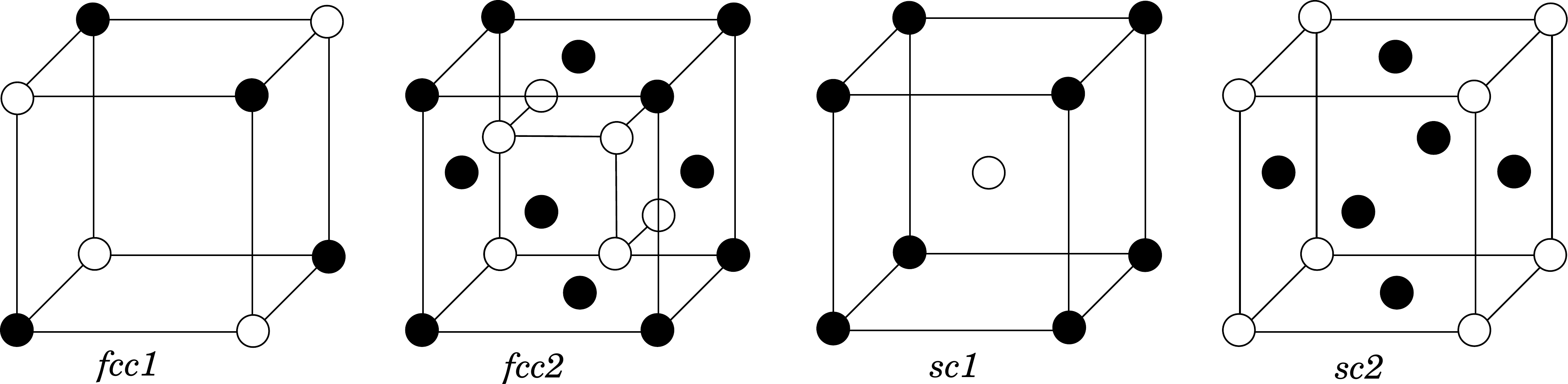}
\end{center}
\vskip -0.5cm
\caption{Binary compounds made of two nuclear species $(A_1,Z_1)$ (black circles) and $(A_2,Z_2)$ (white circles) with face-centered cubic (fcc) and simple cubic (sc) crystal structures. 
}
\label{fig1}
\end{figure*}

\begin{figure*}
\begin{center}
\includegraphics[scale=0.35]{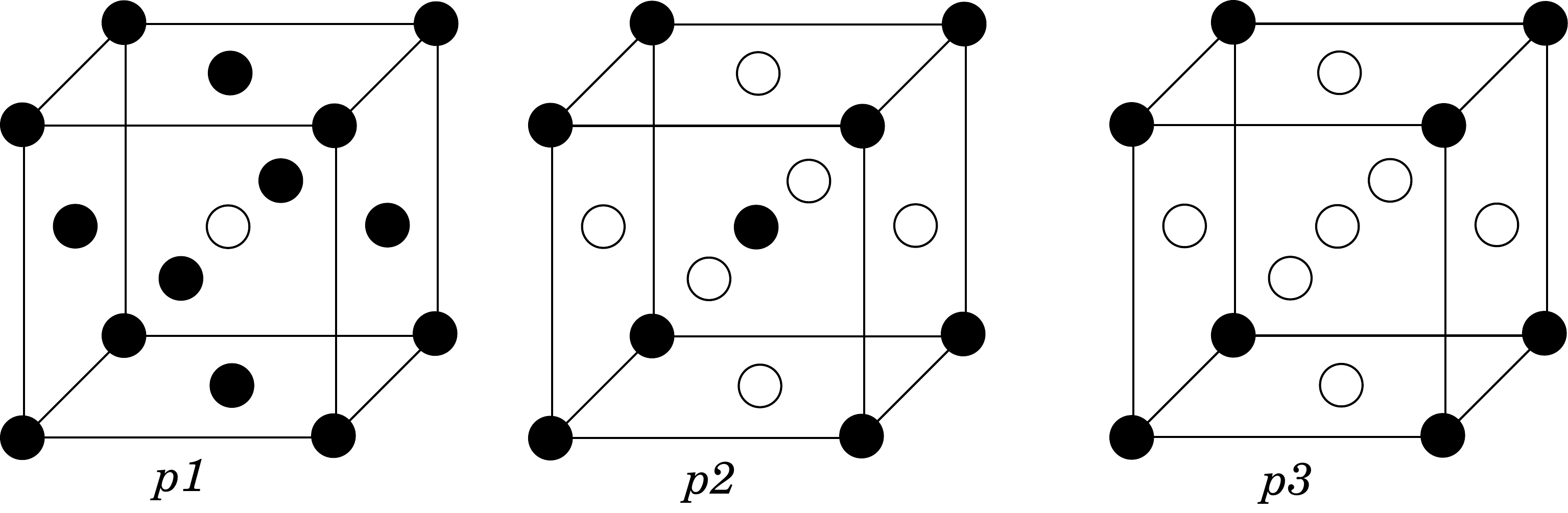}
\end{center}
\vskip -0.5cm
\caption{Binary compounds made of two nuclear species $(A_1,Z_1)$ (black circles) and $(A_2,Z_2)$ (white circles) based on the cubic perovskite 
structure shown in Fig.~\ref{fig:perovskite}. 
}
\label{fig2}
\end{figure*}

\begin{figure*}
\begin{center}
\includegraphics[scale=0.35]{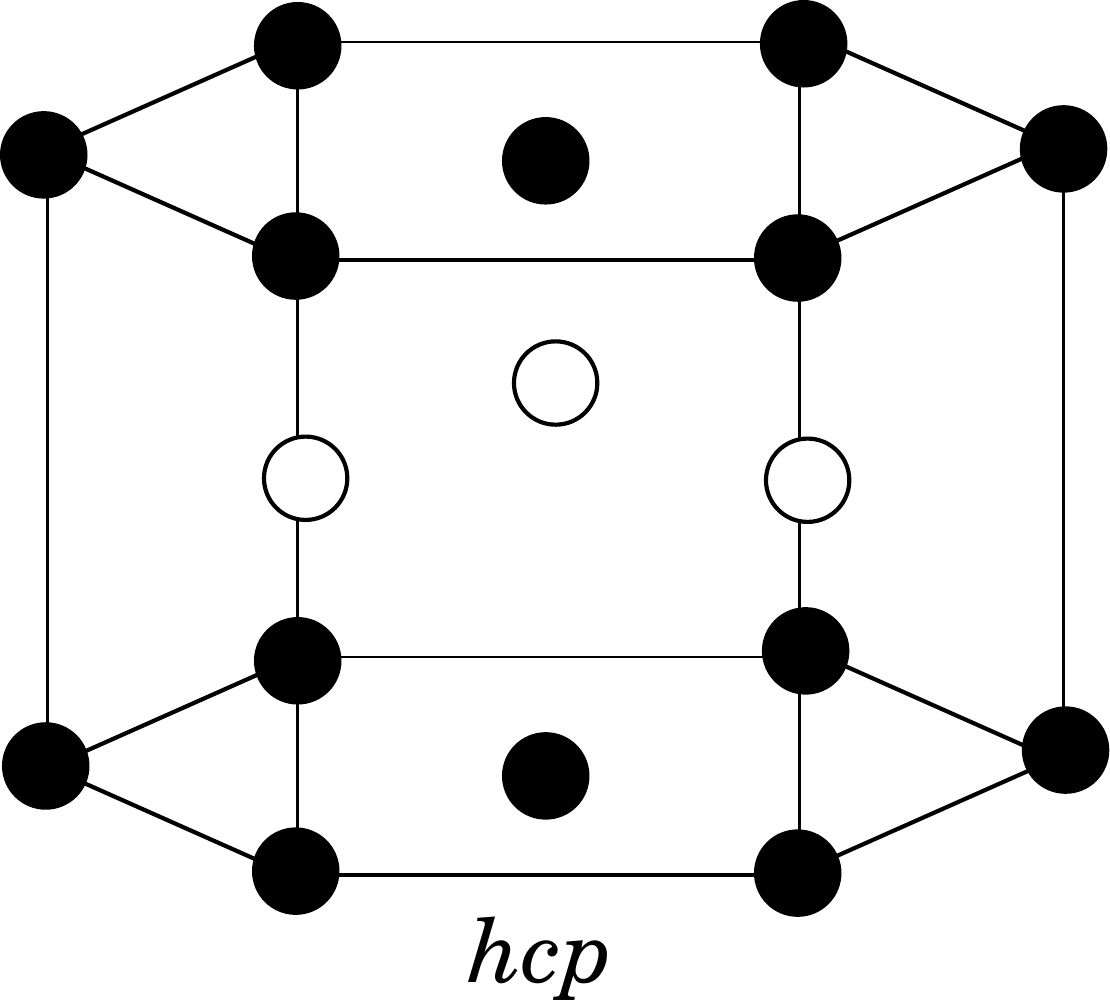}
\end{center}
\vskip -0.5cm
\caption{Binary compound made of two nuclear species $(A_1,Z_1)$ (black circles) and $(A_2,Z_2)$ (white circles) with an hexagonal close-packed (hcp) structure. 
}
\label{fig3}
\end{figure*}

\begin{figure*}
\begin{center}
\includegraphics[scale=0.35]{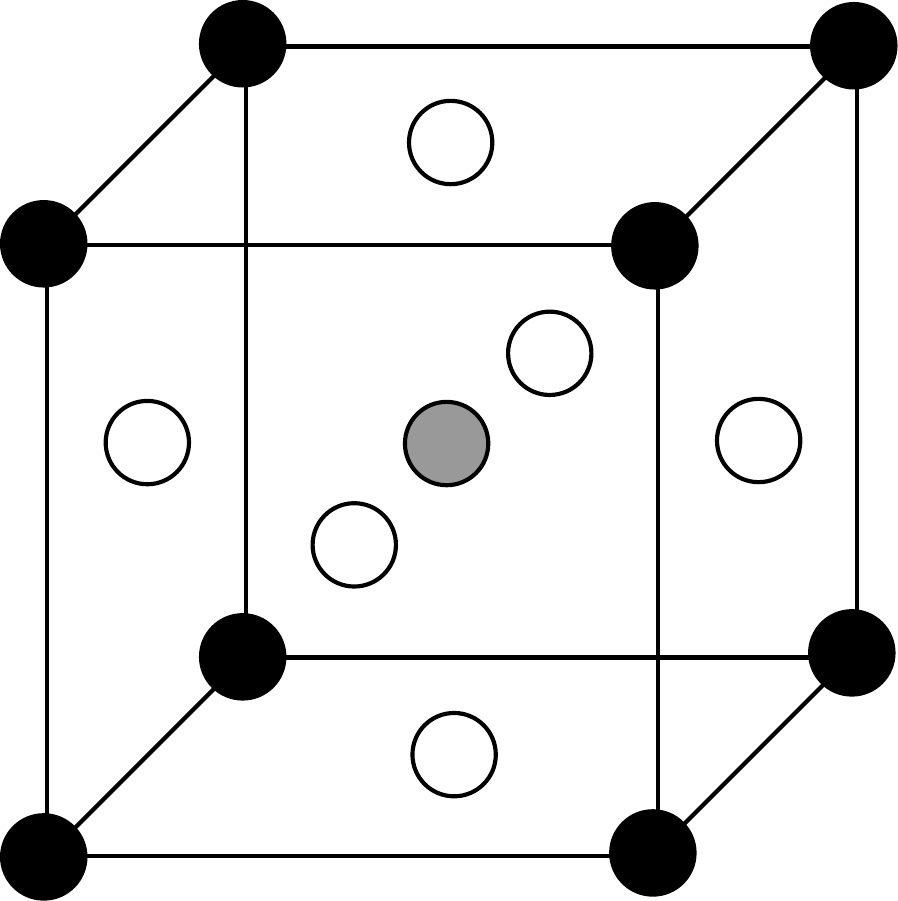}
\end{center}
\vskip -0.5cm
\caption{Ternary compound with cubic perovskite structure made of three nuclear species $(A_1,Z_1)$ (black circles), $(A_2,Z_2)$ (grey circle), and 
$(A_3,Z_3)$ (white circles). 
}
\label{fig:perovskite}
\end{figure*}

The binary compound structures that we consider here are illustrated in Figs.~\ref{fig1}, \ref{fig2} and \ref{fig3}. 
The most familiar example of terrestrial fcc1 compounds is rocksalt - sodium chloride (NaCl). Other such compounds are 
various oxides (e.g. CaO, MgO, NiO, SrO, YbO, ZrO) and carbonitrides (e.g. TiC, TiN, HfC). The prototype of fcc2 compounds is 
fluorite (CaF$_2$). Terrestrial sc1 compounds include for instance cesium chloride (CsCl) and $\beta$-brass (CuZn). Examples 
of terrestrial compounds with sc2 and hcp structures are auricupride (AuCu$_3$) and tungstene carbide (WC)  respectively. 
Kobyakov and Pethick~\cite{kc14} have recently argued that the equilibrium structure of the inner crust of a neutron star 
could be similar to that of baryum titanate (BaTiO$_3$) represented in Fig.~\ref{fig:perovskite}, but made of only one kind 
of nuclear clusters. For this reason, we have also considered binary compounds based on the cubic perovskite structure. The formation 
of ternary compounds will be briefly discussed in Section~\ref{sec:ter-compound}.
Let us stress that stellar compounds differ in two fundamental ways from their terrestrial 
counterparts: first, stellar compounds are made of ``bare'' nuclei; and second, these nuclei are embedded in an essentially 
uniform relativistic electron Fermi gas. 

The structure functions of a binary compound can be generally written as
\begin{equation}\label{eq:f}
f(Z_1,Z_2)  = \bar{Z}^{-4/3}\biggl[\eta Z_1^2 + \zeta Z_2^2+(1-\eta-\zeta)Z_1 Z_2\biggr] \, .
\end{equation}
The numerical values for the lattice constants $C$, $\eta$ and $\zeta$ are indicated in Table~\ref{tab1}. 
Lattice constants of the fcc1, sc1, sc2 and hcp lattices were taken from Ref.~\cite{jog82}. The calculations of the other 
lattice constants can be found in Appendix~\ref{appendix2}. In the limiting case of a pure crystal, the structure function~(\ref{eq:f}) reduces to 
$f(Z,Z)=Z^{2/3}$ independently of $\eta$ and $\zeta$. As can be seen from Table~\ref{tab1}, the sc1 lattice (which coincides in this case with 
a bcc lattice) yields the lowest energy. In particular, pure cubic perovskite structures are unstable in the outer crust of neutron star, but 
might exist in the inner crust due to nuclei-nuclei interactions induced by free neutrons~\cite{kc14}. 

\begin{table}
\begin{tabular}{|c|c|c|c|c|}
\hline
Structure     &  $C$ & $\eta$ & $\zeta$ & $\xi$ \\
\hline
fcc1 & -1.418649 & 0.403981 &  0.403981 & $1/2$\\
fcc2 & -1.39349 & 0.239521 & 0.592901 & $1/3$ \\
sc1 & -1.444231 & 0.389821 & 0.389821 & $1/2$ \\
sc2 & -1.444141 & 0.654710 & 0.154710 & $3/4$ \\
p1 & -1.36588 & 0.785206 & 0.121479 & $4/5$ \\
p2 & -1.36588 & 0.311629 & 0.514083 & $2/5$ \\
p3 & -1.36588 & 0.121479 & 0.660206 & $1/5$ \\
hcp & -1.444083 & 0.345284 &  0.345284 & $1/2$\\
\hline
\end{tabular}
\caption{Structure constants appearing in Eq.~(\ref{eq:f}) for the binary compounds shown in 
Figs.~\ref{fig1}, \ref{fig2}, and \ref{fig3}. The quantity $\xi$ denotes the proportion of nuclei $(A_1,Z_1)$. 
} 
\label{tab1}
\end{table}

\begin{figure*}
\begin{center}
\includegraphics[scale=0.3]{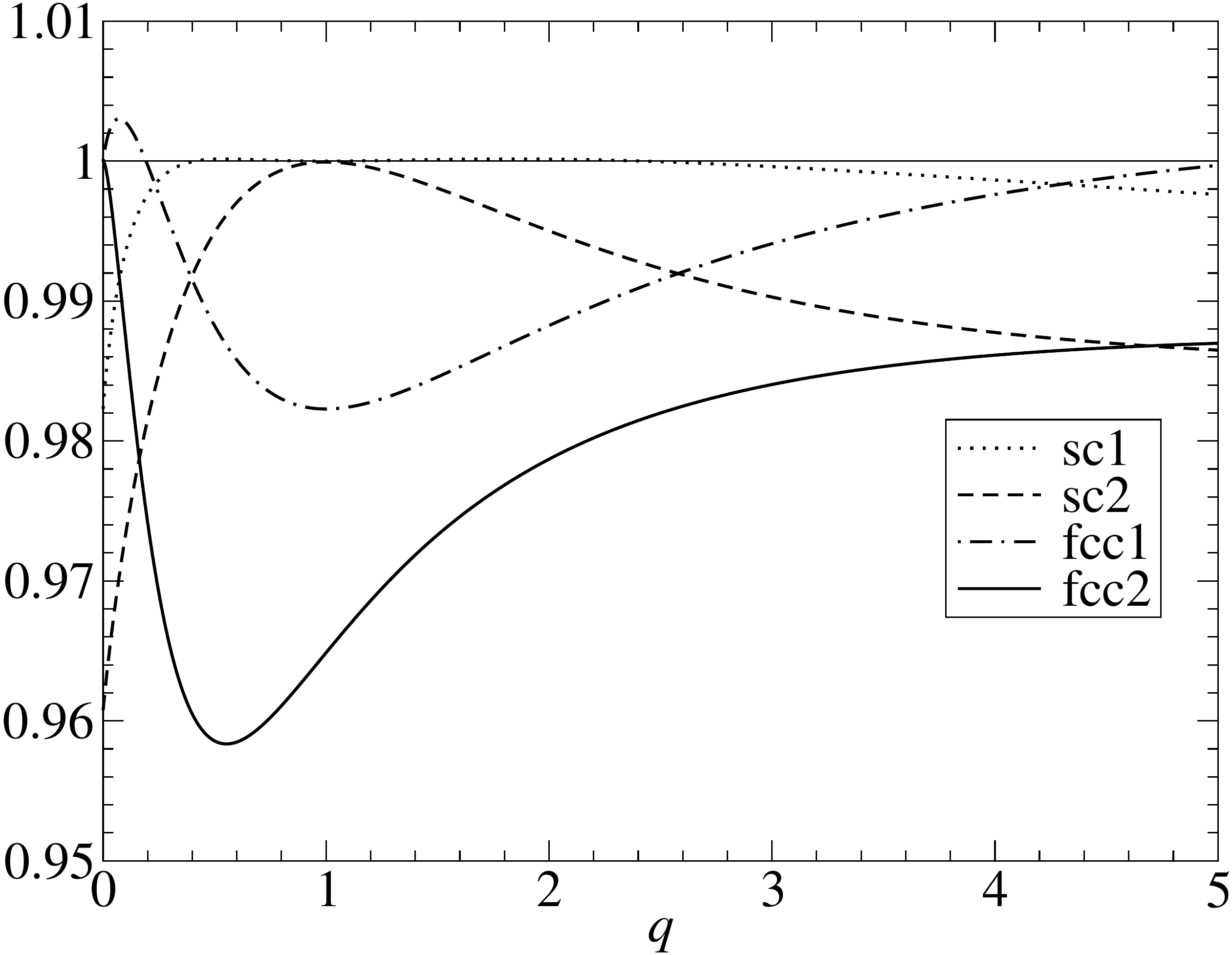}\hskip 0.5cm\includegraphics[scale=0.3]{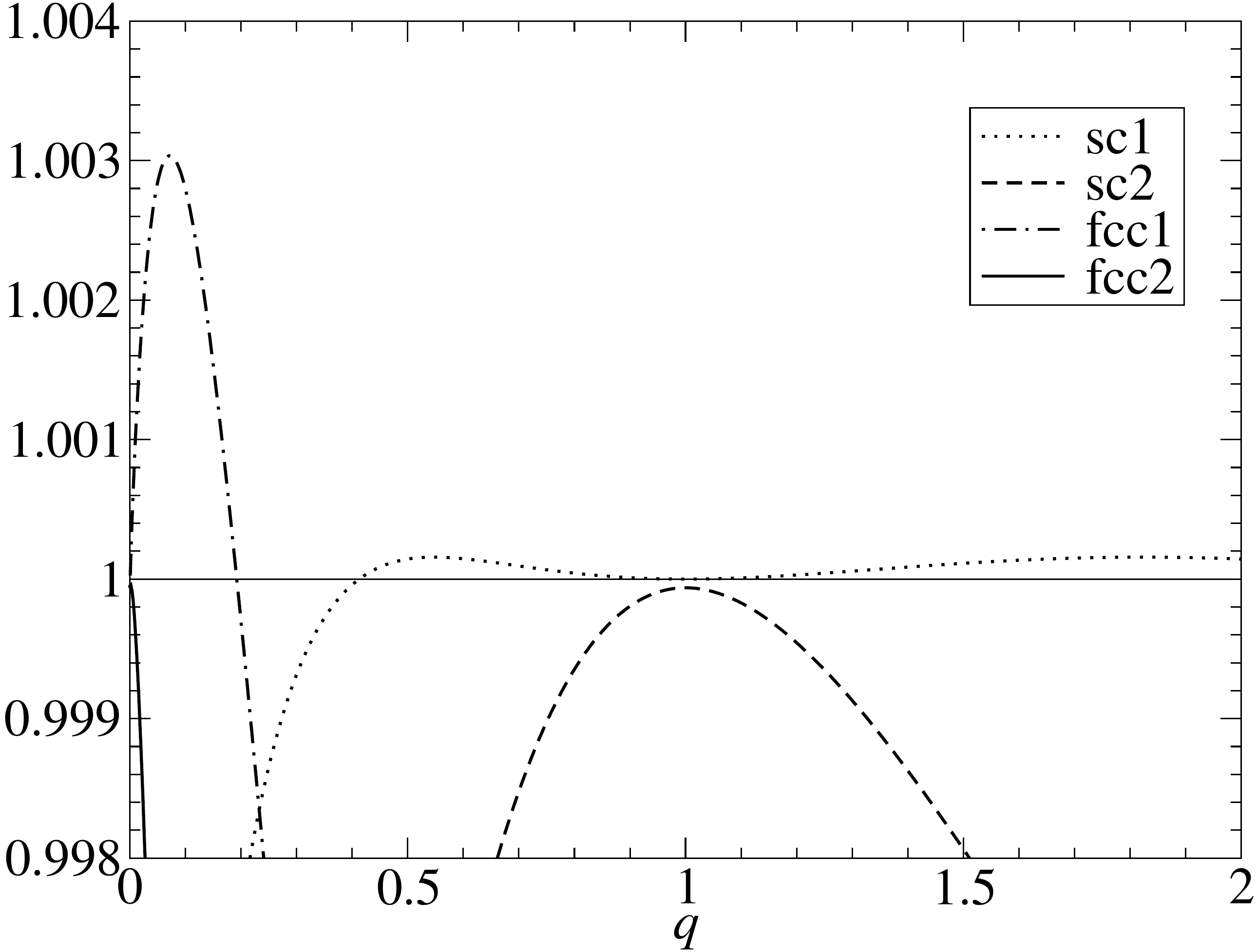}
\end{center}
\vskip -0.5cm
\caption{Values of the dimensionless ratio $\mathcal{R}(q)$ as a function of the charge ratio $q=Z_2/Z_1$ for the binary compounds shown in Figure~\ref{fig1}. 
The thin horizontal line delimits the region of stability against phase separation: a compound is stable if $\mathcal{R}(q)>1$. The right panel shows a close-up view near $q=1$. See text for details. 
}
\label{fig_stab1}
\end{figure*}

\begin{figure*}
\begin{center}
\includegraphics[scale=0.3]{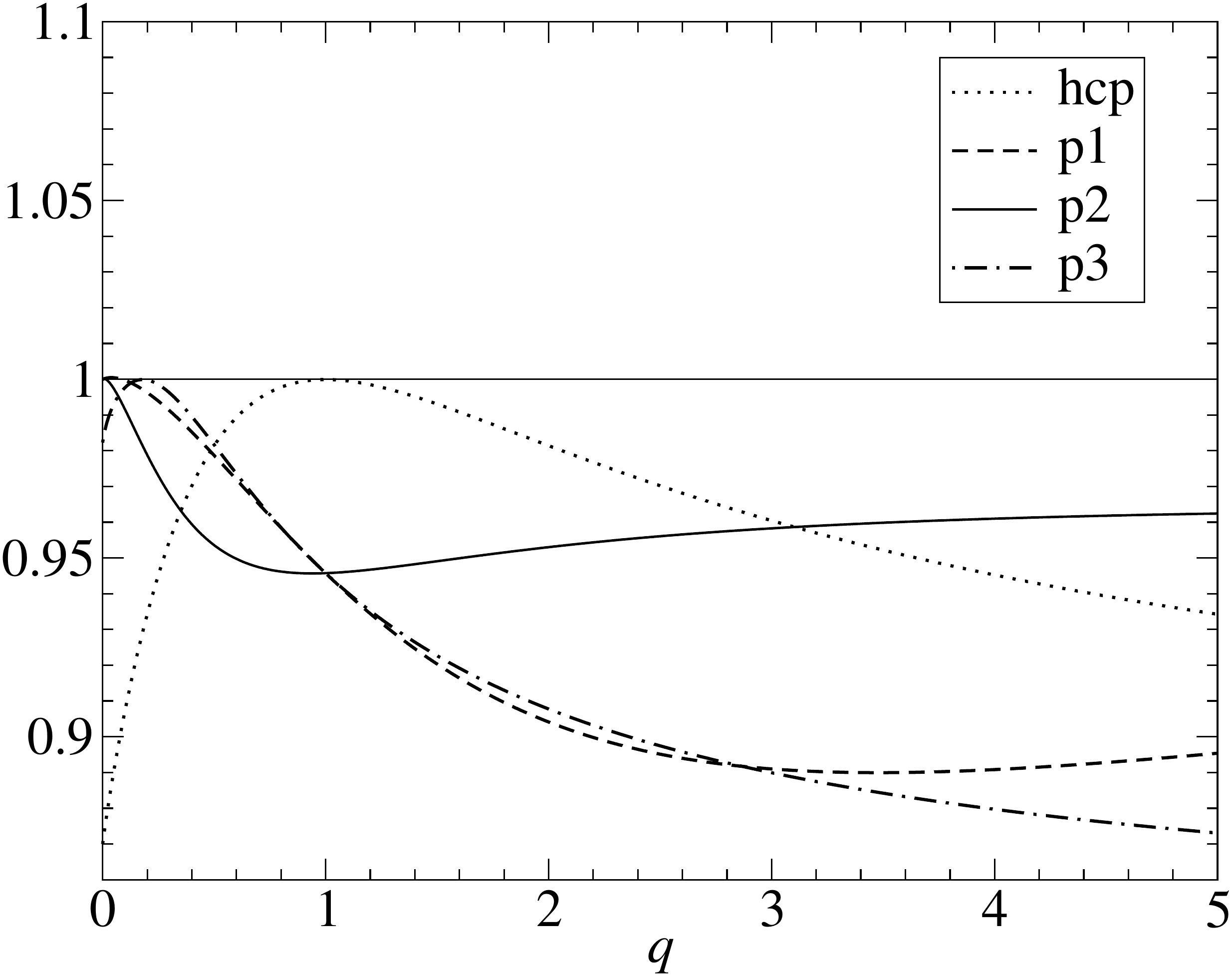}\hskip 0.5cm\includegraphics[scale=0.3]{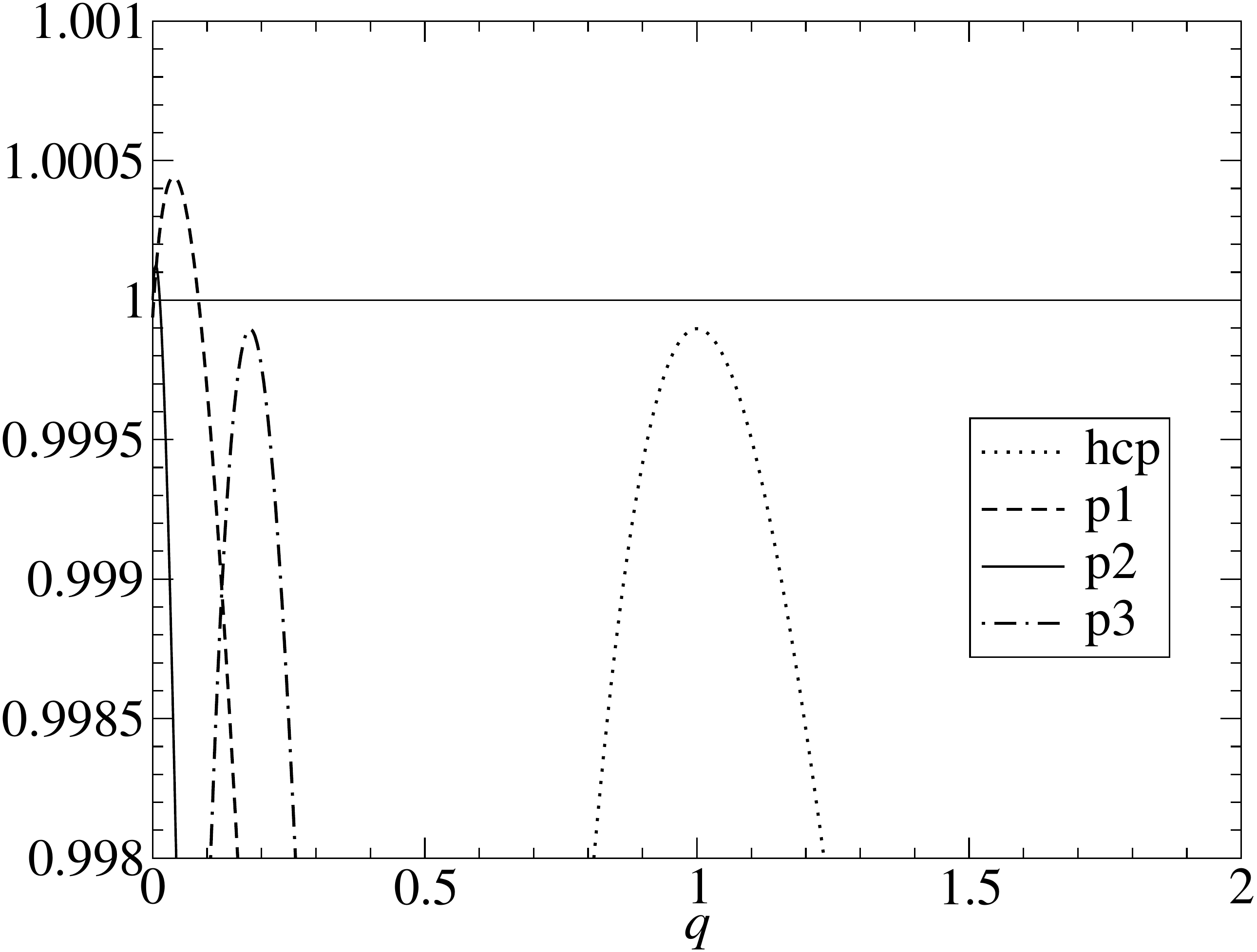}
\end{center}
\vskip -0.5cm
\caption{Same as Figure~\ref{fig_stab1} for the binary compounds shown in Figures~\ref{fig2} and \ref{fig3}. 
}
\label{fig_stab2}
\end{figure*}

As discussed in Section~\ref{sec:compound2}, the stability of a compound is determined by the 
dimensionless ratio 
\begin{equation}
\mathcal{R}(q)=\frac{\tilde{f}(Z_1,Z_2)}{f_{\rm mix}(Z_1,Z_2)}=\frac{C}{C_{\rm bcc}} \frac{\eta +  (1-\eta-\zeta)q +\zeta q^2 }{[\xi+(1-\xi)q]^{1/3}[\xi+(1-\xi)q^{5/3}]}
\end{equation}
where $q\equiv Z_2/Z_1$. The compound is stable if $\mathcal{R}(q)>1$. As shown in Figs.~\ref{fig_stab1} and \ref{fig_stab2}, the sc2, fcc2, p3 and hcp structures do not 
lead to any stable compound since $\mathcal{R}(q) \leq 1$ for any value of $q$. On the other hand, binary compounds with sc1, fcc1, p1, and p2 structures 
can be stable depending on the charge ratios. In particular, the fcc1 structure with a charge ratio $q\simeq 0.07$ yields the most stable compounds with 
$\mathcal{R}\simeq 1.003$, as first pointed out by Dyson~\cite{dyson71}. However, such compounds are not necessarily the most stable ones considering full 
thermodynamic equilibrium with respect to all kinds of weak and strong nuclear reactions~\cite{jog82}. Recalling that the crust of a neutron star is expected to 
be stratified into pure bcc layers with different compositions, binary compounds could be naturally formed at the interfaces by substitution of nuclei, by 
addition of nuclei in the interstices of the bcc lattice, or by both mechanisms. Interstitial compounds with p2 structure are stable against phase separation only 
for very low charge ratios, from $q=0$ to $q\simeq 0.013$ (with $\mathcal{R}$ reaching $1.00013$ at $q\simeq 0.0058$). The p1 type compounds that could be formed from 
both substitution and additions allows for a larger range of values for the charge ratios from $q=0$ to $q\simeq 0.084$ (with $\mathcal{R}$ reaching 
$1.00044$ at $q\simeq 0.039$). Nevertheless, it appears that substitutional compounds with the sc1 structure are the most likely to be present in the crust 
of a neutron star since they are stable against phase separation over a very wide range of values of the charge ratio, from $q\simeq 0.413$ to $q\simeq 2.42$ 
(with $\mathcal{R}$ reaching $1.00016$ at $q\simeq 0.055$ and $q\simeq 1.83$). 
This conclusion is consistent with Monte Carlo simulations of binary ionic mixtures~\cite{oga93}.

\subsection{Stability of cubic perovskite ternary compounds against phase separation}
\label{sec:ter-compound}

In this section, we briefly discuss the possible formation of ternary compounds in the crust of a neutron star. The structure function 
can be quite generally written as  
\begin{equation}\label{eq:fter}
f(Z_1,Z_2,Z_3)  = \bar{Z}^{-4/3}\biggl[\eta_1 Z_1^2 + \eta_{2} Z_2^2+ \eta_3 Z_3^2 + \eta_{12} Z_1 Z_2 + \eta_{13} Z_1 Z_3 + 
   \eta_{23} Z_2 Z_3 \biggr] \, .
\end{equation}
The constants corresponding to the cubic perovskite structure shown in Fig.~\ref{fig:perovskite} are indicated in Table~\ref{tab2}. 
Their calculations can be found in Appendix~\ref{appendix2}.

\begin{table}
\begin{tabular}{|c|c|}
\hline
C & -1.36588 \\
$\eta_1$ &  0.121479 \\
$\eta_2$ &  0.121479 \\
$\eta_3$& 0.514083 \\
$\eta_{12}$& 0.0686701 \\
$\eta_{13}$& 0.149645 \\
$\eta_{23}$& 0.0246441 \\
$\xi_1$ & $1/5$ \\
$\xi_2$ & $1/5$ \\
$\xi_3$ & $3/5$ \\
\hline
\end{tabular}
\caption{Structure constants appearing in Eq.~(\ref{eq:fter}) for the ternary compounds shown in 
Fig.~\ref{fig:perovskite}. The quantities $\xi_1$, $\xi_2$ and $\xi_3$ denote the proportions of 
nuclei $(A_1,Z_1)$,$(A_2,Z_2)$, and $(A_3,Z_3)$ respectively. 
} 
\label{tab2}
\end{table}

\begin{figure*}
\begin{center}
\includegraphics[scale=1]{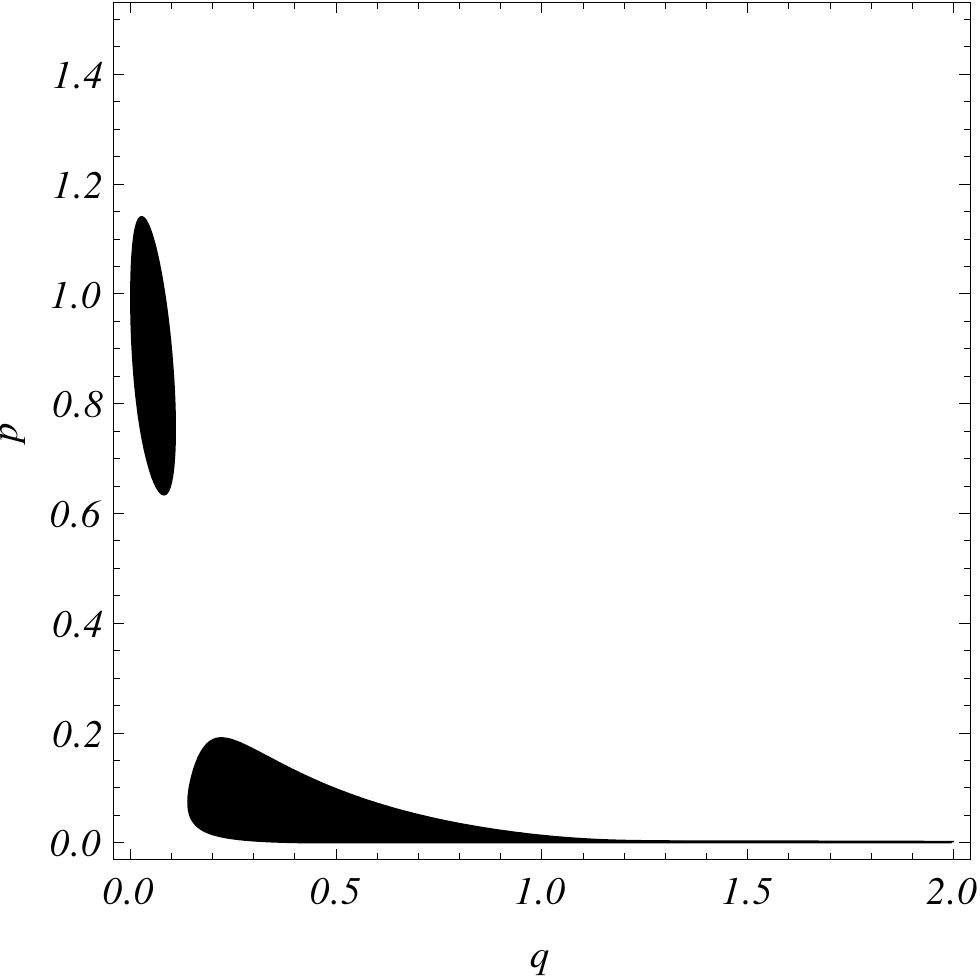}
\end{center}
\vskip -0.5cm
\caption{Stability of cubic perovskite compounds shown in Figure~\ref{fig:perovskite} against phase separation, as a function of the charge ratios 
$q=Z_2/Z_1$ and $p=Z_3/Z_1$. Stable compounds only exist in black regions, where the dimensionless ratio $\mathcal{R}(q,p)$ exceeds unity. 
See text for details. 
}
\label{fig:R-perovskite}
\end{figure*}

As discussed in Section~\ref{sec:compound}, the stability of a ternary compound against phase separation is determined by the dimensionless ratio 
\begin{equation}
\mathcal{R}(q,p)=\frac{\tilde{f}(Z_1,Z_2,Z_3)}{f_{\rm mix}(Z_1,Z_2,Z_3)}
=\frac{C}{C_{\rm bcc}} \frac{\eta_1(1+q^2) + \eta_3 p^2 + \eta_{12} q + \eta_{13} p +\eta_{23} q p }{(\xi_1+\xi_2 q + \xi_3 p)^{1/3}(\xi_1+\xi_2 q^{5/3}+\xi_3 p^{5/3})}\, ,
\end{equation}
where $q\equiv Z_2/Z_1$ and $p\equiv Z_3/Z_1$. The cubic perovskite compound is found to be stable ($\mathcal{R}(q)>1$) in a very restricted domain of 
the charge ratios $q$ and $p$, as shown in Fig.~\ref{fig:R-perovskite}. The maximum is found for $q\simeq 0.0510$ and $p\simeq 0.870$ with $\mathcal{R}\simeq 1.00059$. 
This kind of analysis naturally explains why systems with very different charges are generally more liable to form compounds than systems with similar charges, 
as recently observed in Ref.~\cite{eng16} from systematic phase equilibrium calculations. 

\subsection{Ground-state composition of dense stellar matter}

Determining the ground-state composition of dense matter allowing for fcc1, sc1, p1, and p2 binary compounds as well as the cubic perovskite ternary compounds 
still remains computationally very expensive since at each pressure $P$ the Gibbs free energy must be calculated for all possible combinations of nuclei 
($\sim 7\times 10^7$ for binary compounds, $\sim 6\times 10^{11}$ for ternary compounds). For this reason, we shall proceed as follows. 

First, the equilibrium composition of the crust is determined by considering only pure phases (for which the most stable crystal structure is the bcc lattice). 
To this end, for each pressure $P$, we minimize the Gibbs free energy per nucleon $g(A,Z,P)$ over all possible nuclei $(A,Z)$. We have made use of experimental 
atomic masses from the 2012 Atomic Mass Evaluation~\cite{audi12} supplemented with the Brussels-Montreal microscopic nuclear mass model HFB-24~\cite{goriely2013} 
for the masses that have not yet been measured. Starting from the initial value $P=3\times 10^{-11}$~MeV~fm$^{-3}$, we have increased the pressure with a step 
$\Delta P = 0.003 P$ until the onset of neutron dripping out of nuclei, which delimits the boundary between the outer and inner regions of the crust.  
Results are summarized in Table~\ref{tab:hfb24-pure}. 
Let us consider that the equilibrium nucleus thus found is $(A,Z)$ in the range of pressures $P_{\rm min}\leq P\leq P_{\rm max}$. The compounds most likely 
to appear in this region are those yielding the lowest values for the threshold pressure $P_{\rm thres}$, which is completely determined by 
Eq.~(\ref{eq:threshold-condition}) and can therefore be tabulated once and for all. In the ultrarelativistic regime, $P_{\rm thres}$ can be 
accurately estimated from Eq.~(\ref{eq:threshold-pressure}). Some compounds may yield values for the threshold electron Fermi energy such that 
$\mu_e^{\rm thres} < m_e c^2$ ; such transitions are energetically forbidden and must therefore be ignored. A compound will be energetically favored 
if the corresponding threshold pressure $P_{\rm thres}$ lies in the range $P_{\rm min}\leq P_{\rm thres}\leq P_{\rm max}$. The analysis of the mass tables 
suggests that the compounds most likely to exist in the outer crust of a neutron star are those made of nuclei from neighboring layers, as previously found in 
Ref.~\cite{jog82}. By inspecting Table~\ref{tab:hfb24-pure}, it can be seen that the corresponding charge ratios vary from about $q=0.93$ for $^{62}$Ni+$^{58}$Fe to 
$1.5$ for $^{80}$Ni+$^{124}$Mo. Therefore, only binary compounds with sc1 structure need to be considered, as can be inferred from Fig.~\ref{fig_stab1}  
(see also Refs.~\cite{kozh2012,kozh2015}). 

\begin{figure}
 \centering
 \includegraphics[scale=.5]{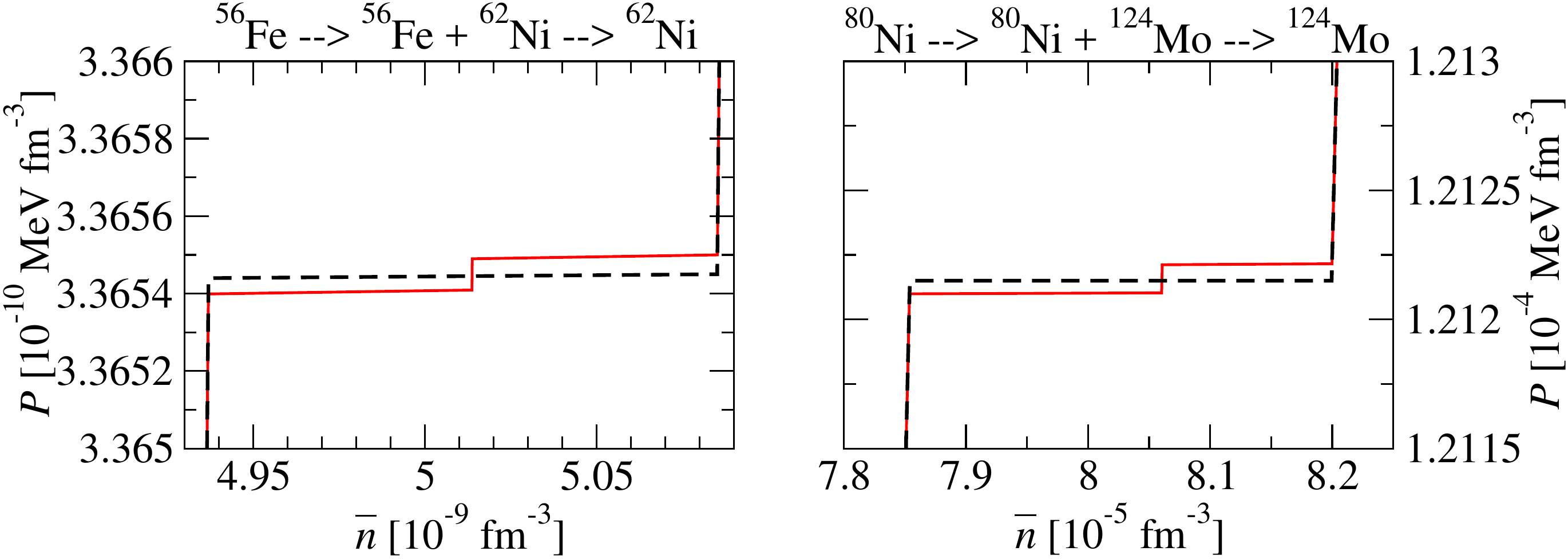}
 \caption{Pressure versus mean baryon number density in the outer crust of a cold nonaccreting neutron star. Results were obtained using experimental masses from the 2012 Atomic Mass Evaluation~\cite{audi12} supplemented with the Brussels-Montreal nuclear mass model HFB-24~\cite{goriely2013}. Two regions of the crust are highlighted corresponding to the change of composition from $^{56}$Fe to $^{62}$Ni (left panel), and from $^{80}$Ni to $^{124}$Mo (right panel). 
 In both cases, a binary compound with the sc1 structure shown in Fig.~\ref{fig1} is present in the intermediate layers. The dashed lines represent the transition considering pure crystalline phases only. 
 } 
 \label{fig:eos-hfb24-zoom}
\end{figure}

For each pair $(A_1,Z_1)$ and $(A_2,Z_2)$ of adjacent nuclei shown in Table~\ref{tab:hfb24-pure}, we have solved numerically Eq.~(\ref{eq:gibbs-transition-general}) with 
$(A,Z)=(A_1,Z_1)$ and $(A,Z)=(A_2,Z_2)$. In this way, we have determined the threshold pressures $P_{1\rightarrow1+2}$ and $P_{1+2\rightarrow2}$, as well as the densities 
for the appearance and disappearance of the different phases without any further approximation. Results are summarized in Table~\ref{tab:hfb24-compound}. We have studied 
the stability of these compounds against phase separation. We have found that the direct comparison of the Gibbs free energies per nucleon is equivalent to the simple 
criterion~(\ref{eq:separation}) obtained by expanding $g$ to first order in $\alpha$. Contrary to the results obtained in Ref.~\cite{jog82}, we do not find any stable 
compound composed of different isotopes, as anticipated in Section~\ref{sec:compound2}. Using the numerical results, we have tested the precision of the analytical 
formulas for the threshold densities and pressures. The errors amount at most to $0.2$\% for the pressures and $0.07$\% for the densities. The approximate expressions 
obtained under the assumption of ultrarelativistic electrons are less reliable. However, their precision is expected to increase with increasing $\mu_e^{1\rightarrow2}$ 
thereby with increasing depth below the stellar surface (see Table~\ref{tab:hfb24-pure}). Leaving aside the shallowest layer containing 
the compound made of $^{56}$Fe+$^{62}$Ni (the associated threshold electron Fermi energy is less than twice the electron rest mass energy, see Table~\ref{tab:hfb24-pure}), 
the errors amount at most to about $16$\% for the pressures and $8$\% for the densities. The errors on the pressures and on the densities are the largest for the 
compound made of $^{62}$Ni+$^{58}$Fe, and drop to about $0.2$\% and $0.08$\% respectively for the deepest compound made of $^{121}$Y+$^{120}$Sr. As illustrated in 
Fig.~\ref{fig:eos-hfb24-zoom}, the impact of binary compounds on the equation of state of the outer crust of a nonaccreting neutron star is very small.

\begin{table}
\caption{Composition of the outer crust of a cold nonaccreting neutron star considering only pure body-centered cubic crystals made of nuclei with atomic number $Z$ and 
mass number $A$. Results were obtained using experimental masses from the 2012 Atomic Mass Evaluation~\cite{audi12} supplemented with the Brussels-Montreal nuclear mass model HFB-24~\cite{goriely2013}. The mean baryon number densities $\bar n$ are measured in units of fm$^{-3}$, 
the transition pressures $P_{1\rightarrow2}$ are in units of MeV~fm$^{-3}$, and the threshold electron Fermi energies $\mu_e^{1\rightarrow2}$ are in units of MeV. See text for details.} 
\label{tab:hfb24-pure}
\begin{tabular}{cccccc}
\hline 
 $Z$ & $A$ & $\bar n_{\rm min}$   & $\bar n_{\rm max}$ & $P_{1\rightarrow2}$ & $\mu_e^{1\rightarrow 2}$\\
\hline 
 26 & 56 & $-$                       & $4.94 \times 10^{-9}$ & $3.36 \times 10^{-10}$ & $0.96$ \\
 28 & 62 & $5.09 \times 10^{-9}$  & $1.59 \times 10^{-7}$ & $4.20 \times 10^{-8}$ & $1.97$ \\
 26 & 58 & $1.60 \times 10^{-7}$  & $1.65 \times 10^{-7}$ & $4.39 \times 10^{-8}$ & $2.67$ \\
 28 & 64 & $1.70 \times 10^{-7}$  & $8.01 \times 10^{-7}$ & $3.56 \times 10^{-7}$ & $4.15$ \\
 28 & 66 & $8.28 \times 10^{-7}$  & $9.21 \times 10^{-7}$ & $4.12 \times 10^{-7}$ & $6.21$ \\
 36 & 86 & $9.42 \times 10^{-7}$  & $1.86 \times 10^{-6}$ & $1.03 \times 10^{-6}$ & $5.13$ \\
 34 & 84 & $1.92 \times 10^{-6}$  & $6.79 \times 10^{-6}$ & $5.57 \times 10^{-6}$ & $7.83$ \\
 32 & 82 & $7.04 \times 10^{-6}$  & $1.67 \times 10^{-5}$ & $1.77 \times 10^{-5}$ & $10.49$ \\
 30 & 80 & $1.74 \times 10^{-5}$  & $3.46 \times 10^{-5}$ & $4.44 \times 10^{-5}$ & $13.25$ \\
 28 & 78 & $3.62 \times 10^{-5}$  & $6.64 \times 10^{-5}$ & $1.00 \times 10^{-4}$ & $16.85$ \\
 28 & 80 & $6.83 \times 10^{-5}$  & $7.85 \times 10^{-5}$ & $1.21 \times 10^{-4}$ & $23.06$ \\
 42 & 124 & $8.21 \times 10^{-5}$ & $1.21 \times 10^{-4}$ & $2.05 \times 10^{-4}$ & $19.09$ \\
 40 & 122 & $1.26 \times 10^{-4}$ & $1.56 \times 10^{-4}$ & $2.75 \times 10^{-4}$ & $20.62$ \\
 39 & 121 & $1.59 \times 10^{-4}$ & $1.63 \times 10^{-4}$ & $2.85 \times 10^{-4}$ & $20.84$ \\
 38 & 120 & $1.67 \times 10^{-4}$ & $1.95 \times 10^{-4}$ & $3.54 \times 10^{-4}$ & $22.92$ \\
 38 & 122 & $1.99 \times 10^{-4}$ & $2.40 \times 10^{-4}$ & $4.55 \times 10^{-4}$ & $24.40$ \\
 38 & 124 & $2.44 \times 10^{-4}$ & $2.56 \times 10^{-4}$ & $4.87 \times 10^{-4}$ & $-$ \\

\hline
\end{tabular}
\end{table}

\begin{table}
\caption{Composition of the outer crust of a cold nonaccreting neutron star allowing for binary compounds with the sc1 structure shown in Fig.~\ref{fig1}. 
Results were obtained using experimental masses from the 2012 Atomic Mass Evaluation~\cite{audi12} supplemented with the Brussels-Montreal nuclear mass model HFB-24~\cite{goriely2013}. 
The columns give the atomic and mass numbers of the two nuclei, the maximum mean baryon number density $\bar n_1^{\rm max}$ at which the pure crystal $(A_1,Z_1)$ is present, the transition pressure $P_{1\rightarrow1+2}$ for the formation of the compound, the lowest and highest densities (respectively $\bar n_{1+2}^{\rm min}$ and $\bar n_{1+2}^{\rm max}$) at which the compound exists, the threshold pressure $P_{1+2\rightarrow2}$ for the disappearance of the compounds, and the minimum density $\bar n_{2}^{\rm min}$ at which the pure crystal $(A_2,Z_2)$ appears. The densities are measured in units of fm$^{-3}$, and the pressures are in units of MeV~fm$^{-3}$.}
\label{tab:hfb24-compound}
% \footnotesize
\scriptsize
\begin{tabular}{cccccccccc}
\hline 
 $Z_1$ & $A_1$ & $Z_2$ & $A_2$ & $\bar n_1^{\rm max}$ & $P_{1\rightarrow1+2}$ & $\bar n_{1+2}^{\rm min}$ &  $\bar n_{1+2}^{\rm max}$ & $P_{1+2\rightarrow2}$  & $\bar n_{2}^{\rm min}$  \\
\hline 
26 & 56 & 28 & 62 & $4.93698 \times 10^{-9}$ & $3.36539 \times 10^{-10}$ & $5.01371 \times 10^{-9}$ & $5.01376 \times 10^{-9}$ & $3.36544 \times 10^{-10}$ & $5.08514 \times 10^{-9}$\\
28 & 62 & 26 & 58 & $1.59126 \times 10^{-7}$ & $4.20338 \times 10^{-8}$ & $1.59572 \times 10^{-7}$ & $1.59591 \times 10^{-7}$ & $4.20407\times 10^{-8}$ & $1.60071\times 10^{-7}$ \\
26 & 58 & 28 & 64 & $1.65281 \times 10^{-7}$ & $4.39084 \times 10^{-8}$ & $1.67532 \times 10^{-7}$ & $1.67537 \times 10^{-7}$ & $4.39100 \times 10^{-8}$ & $1.69630 \times 10^{-7}$\\
28 & 66 & 36 & 86 & $9.23299 \times 10^{-7}$ & $4.13609 \times 10^{-7}$ & $9.33458 \times 10^{-7}$ & $9.33710 \times 10^{-7}$ & $4.13758 \times 10^{-7}$ & $9.41661 \times 10^{-7}$ \\
36 & 86 & 34 & 84 & $1.85904 \times 10^{-6}$ & $1.02956 \times 10^{-6}$ & $1.88857 \times 10^{-6}$ & $1.88859 \times 10^{-6}$ & $1.02957 \times 10^{-6}$ & $1.91982 \times 10^{-6}$\\
34 & 84 & 32 & 82 & $6.79092 \times 10^{-6}$ & $5.57552 \times 10^{-6}$ & $6.90842 \times 10^{-6}$ & $6.90850 \times 10^{-6}$ & $5.5756 \times 10^{-6}$ & $7.03316 \times 10^{-6}$ \\
32 & 82 & 30 & 80 & $1.66930 \times 10^{-5}$ & $1.76809 \times 10^{-5}$ & $1.70089 \times 10^{-5}$ & $1.70091 \times 10^{-5}$ &  $1.76812 \times 10^{-5}$ &$1.73455 \times 10^{-5}$ \\
30 & 80 & 28 & 78 & $3.45867 \times 10^{-5}$ & $4.44090 \times 10^{-5}$ & $3.53057 \times 10^{-5}$ & $3.53062 \times 10^{-5}$ & $4.44098 \times 10^{-5}$ &$3.60754 \times 10^{-5}$\\
28 & 80 & 42 & 124 & $7.85386 \times 10^{-5}$ & $1.21210 \times 10^{-4}$ & $8.06025 \times 10^{-5}$ & $8.06081 \times 10^{-5}$ & $1.21221 \times 10^{-4}$ & $8.19983 \times 10^{-5}$\\
42 & 124 & 40 & 122 & $1.21516 \times 10^{-4}$ & $2.04847 \times 10^{-4}$ & $1.23392 \times 10^{-4}$ & $1.23392 \times 10^{-4}$ & $2.04849 \times 10^{-4}$ &$1.25360 \times 10^{-4}$\\
40 & 122 & 39 & 121 & $1.56784 \times 10^{-4}$ & $2.76056 \times 10^{-4}$ & $1.58063 \times 10^{-4}$ & $1.58064 \times 10^{-4}$ &  $2.76058 \times 10^{-4}$ &  $1.59375 \times 10^{-4}$\\
39 & 121 & 38 & 120 & $1.63675 \times 10^{-4}$ & $2.86036 \times 10^{-4}$ & $1.65058 \times 10^{-4}$ & $1.65058 \times 10^{-4}$ &  $2.86037 \times 10^{-4}$ &  $1.66476 \times 10^{-4}$ \\
\hline
\end{tabular}
\normalsize
\end{table}

\section{Conclusions}

It has been generally thought that the outer crust of a cold nonaccreting neutron star consists of different layers, each of which 
are made of a pure body-centered cubic ionic crystal in a charge compensating background of highly degenerate electrons. We have analyzed 
the stability of such layer against its conversion (due to weak and strong nuclear processes) into a multinary ionic compound with an 
arbitrary composition. We have derived general analytical formulas for the pressure at the onset of the transition, as well as for 
the densities of the different phases irrespective of the degree of relativity of the electron gas. These expressions take particularly 
simple forms in the limit of ultrarelativistic electrons, see Eqs.~(\ref{eq:threshold-pressure}), (\ref{eq:threshold-density}) and 
(\ref{eq:threshold-density-min}) respectively. 

A necessary condition for the formation of a compound is that it must be stable against the separation into pure coexisting phases. We have shown 
that the stability condition, embedded in Eq.~(\ref{eq:separation}), is uniquely determined by the compound structure and composition irrespective 
of the stellar conditions. In particular, we have thus shown that the stability of a compound against phase separation in dense matter 
depends neither on the pressure nor on the degree of relativity of the electron gas, as recently noticed in Ref.~\cite{eng16} from a systematic 
search of equilibrium phases. Moreover, with this simple criterion, it can be easily shown that systems with very different charges are more likely to 
form stable compounds than systems with similar charges, as observed in Ref.~\cite{eng16}. 

However, equilibrium with respect to weak and strong nuclear interactions imposes very stringent constraints on the nuclear species present 
in the crust of a neutron star. Typically, 
the charge numbers of equilibrium nuclides are $Z\sim 30-40$. For this reason, the formation of disordered compounds and multinary compounds 
made of a large variety of different nuclei appears very unlikely (see, e.g., Ref.~\cite{iga01}). On the other hand, ordered binary ionic 
compounds are generally present at the interface between two pure adjacent crustal layers. Their existence is uniquely determined by their 
stability against phase separation, and can thus be very easily assessed. In particular, compounds made of different isotopes are unstable. 
By examining different cubic and noncubic lattices, we have found that substitutional compounds having the same structure as cesium chloride 
are the most likely to be formed in the outer crust of a nonaccreting neutron star, whereas compounds similar to auricupride (AuCu$_3$) or 
tungstene carbide (WC) are all unstable irrespective of their composition. Likewise, the formation of ternary compounds with cubic perovskite 
structure such as baryum titanate (BaTiO$_3$) is found to be highly improbable. 

Using experimental atomic mass data from the 2012 Atomic Mass Evaluation~\cite{audi12} supplemented with the microscopic HFB-24 nuclear mass model~\cite{goriely2013}, 
we have calculated the ground-state structure and the equation of state of the outer crust of a cold nonaccreting neutron star allowing for 
binary compounds. These calculations have confirmed that compounds with cesium chloride structure can be formed at the boundary between 
pure crustal layers. With these numerical results, we have also been able to assess the precision of our analytical formulas: the errors never 
exceed $0.2\%$ for the threshold pressures, and $0.07\%$ for the densities of the solid phases. Although the impact of binary compounds on the 
equation of state has been found to be very small, their presence may have important implications for the thermal and mechanical properties 
(especially the brittleness) of the crust (see, e.g. Refs.~\cite{kozh2012,kozh2015,iga03}). In our investigation, we have neglected electron 
exchange and polarization effects, as well as quantum zero point motion of ions about their equilibrium positions. Although these corrections 
are very small~\cite{dyson71,witten74}, they may affect the stability of ionic compounds and thus need to be closely examined. 

From our analysis, we expect a much large variety of ordered and disordered multinary ionic compounds to form in the core of white dwarfs, 
and in the crust of accreting neutron stars (see, e.g. Refs.~\cite{eng16,horo09}). This warrants further studies.

\appendix
\section{On transitions to a multicomponent solid phase such that $\bar Z/\bar A=Z/A$}
\label{appendix1}
Let us consider the transition from of a solid made of only one type of nuclei $(A,Z)$ to a multi-component solid made of nuclei $(A_i,Z_i)$ 
such that $\bar Z/\bar A=Z/A$. Solving the threshold condition~(\ref{eq:gibbs-transition-general}) after expanding the Gibbs free energies per nucleon 
to first order in $\alpha$ leads to the following expression for the maximum mean nucleon number density of the pure solid phase 
\begin{equation}\label{eq:exact-threshold-density-peculiar}
\bar n^{\rm max} = \biggl[\frac{\bar M^\prime}{\bar A  m_e}-\frac{M^\prime(A,Z)}{A m_e}\biggr]^3 \frac{(A/Z)^4}{(\lambda_e C_{\rm bcc} \alpha)^3}
\biggl[ Z^{2/3} - \tilde{f}(\{Z_i\})\biggr]^{-3}\, .
\end{equation}
Because $\alpha\ll 1$, the threshold density for the onset of such transitions is thus likely to lie well above the threshold densities for any transitions 
accompanied by a discontinous change of proton fraction ($\bar Z/\bar A\neq Z/A$). 
As a matter of fact, $\bar n^{\rm max}$ diverges as $\alpha \rightarrow 0$ for transitions such that $\bar Z/\bar A=Z/A$ whereas $\bar n^{\rm max}$ 
remains finite for any other transitions, as can be seen by comparing Eqs.~(\ref{eq:exact-threshold-density-peculiar}) and 
(\ref{eq:exact-threshold-density}), with 
\begin{equation}
 x_r=\gamma_e^{\rm thres} \sqrt{1-\frac{1}{(\gamma_e^{\rm thres})^2}} 
\end{equation}
using Eq.~(\ref{eq:exact-xr}).

\section{Structure function of cubic binary compounds}
\label{appendix2}

The structure function $f(Z_1,Z_2)$ of the fluorite (fcc2) and cubic perovskite (p1, p2, and p3) lattices shown in Figs.~\ref{fig1} 
and \ref{fig2} can be determined from the calculations of Refs.~\cite{bald92,jog82} using the lattice constants $C_{\rm sc}$, $C_{\rm bcc}$, and 
$C_{\rm fcc}$ of the pure sc, bcc, and fcc lattices, respectively. For these latter constants, we shall use the values given in 
Ref.~\cite{jog82} and indicated in Table~\ref{tab1}: considering that if the two nuclear species are the same, the fcc1 lattice reduces to a sc lattice, 
the sc1 lattice to a bcc lattice, and the sc2 lattice to a fcc lattice, we thus have $C_{\rm sc}=C_{\rm fcc1}$, $C_{\rm bcc}=C_{\rm sc1}$, 
and $C_{\rm fcc}=C_{\rm sc2}$. The calculations of the lattice constants of the fluorite and cubic perovskite structures are 
presented in the following sections.

\subsection{Fluorite} 

From dimensional analysis, the lattice energy density can be quite generally expressed as 
\begin{equation}
 \mathcal{E}_L = n_N \frac{e^2}{a_{N}}\left( c_1 Z_1^2 + c_2 Z_2^2 + c_{12} Z_1 Z_2\right)\, , 
\end{equation}
where $n_N$ denotes the mean number density of nuclei $(A_1,Z_1)$ and $(A_2,Z_2)$, whereas $a_N=(4\pi n_N/3)^{-1/3}$ is 
the ion-sphere radius. Using the electric charge neutrality condition $n_e=\bar Z n_N$, the lattice energy density 
can be equivalently written in the form of Eq.~(\ref{eq:EL}) with the structure function given by Eq.~(\ref{eq:f}). The 
corresponding constants are given by 
\begin{equation}
 C=\left(\frac{4\pi}{3}\right)^{1/3}(c_1+c_2+c_{12})\, ,
\end{equation}
\begin{equation}
 \eta=\frac{c_1}{c_1+c_2+c_{12}}\, , 
\end{equation}
\begin{equation}
 \zeta=\frac{c_2}{c_1+c_2+c_{12}}\, .
\end{equation}
These constants can be determined by considering limiting cases as follows. 
\begin{itemize}
 \item $Z_1=0$

The nuclei $(A_2,Z_2)$ form a sc sublattice, therefore the lattice energy $\mathcal{E}_L$ must coincide with 
that of a sc lattice of nuclei $(A_2,Z_2)$, i.e. 
\begin{equation}
n_N \frac{e^2}{a_{N}}c_2 Z_2^2 = n_2 \frac{e^2}{a_2} \left(\frac{3}{4\pi}\right)^{1/3} C_{\rm sc} Z_2^2\, , 
\end{equation}
where $a_2=(4\pi n_2/3)^{-1/3}$, and $n_2=(2/3)n_N$, as can be easily seen from Fig.~\ref{fig1}. We thus obtain 
\begin{equation}
c_2 = \left(\frac{3}{4\pi}\right)^{1/3} \left(\frac{2}{3}\right)^{4/3} C_{\rm sc} \, . 
\end{equation}

 \item $Z_2=0$
 
The nuclei $(A_1,Z_1)$ form a fcc sublattice. Following the same reasoning as above, we find 
\begin{equation}
c_1=  \left(\frac{3}{4\pi}\right)^{1/3}\left(\frac{1}{3}\right)^{4/3} C_{\rm fcc} \, . 
\end{equation}

\item $Z_1=Z_2=1$ 

The lattice energy density reduces to 
\begin{equation}
 \mathcal{E}_L = n_N \frac{e^2}{a_{N}}\left( c_1  + c_2 + c_{12}\right)\, , 
\end{equation}
which can be directly compared to the expression obtained in Ref.~\cite{bald92}: 
\begin{equation}
 \mathcal{E}_L = -1.728906\, n_N \frac{e^2}{2 a_{N}}\, .  
\end{equation}
We can thus determine the remaining coefficient $c_{12}$ from the equation 
\begin{equation}
2\left( c_1  + c_2 + c_{12}\right)= -1.728906\, . 
\end{equation}

\end{itemize}

\subsection{Cubic perovskites}

From dimensional analysis, the lattice energy density of a ternary compound can be quite generally expressed as 
\begin{equation}
 \mathcal{E}_L = n_N \frac{e^2}{a_{N}}\left( c_1 Z_1^2 + c_2 Z_2^2 + c_3 Z_3^2 + c_{12} Z_1 Z_2 + c_{13} Z_1 Z_3 + c_{23} Z_2 Z_3\right)\, . 
\end{equation}
Alternatively, the lattice energy density can be written in the form~(\ref{eq:EL}) with the structure function (\ref{eq:fter}). 
The different coefficients are related to each by the following equations: 
\begin{equation}
 C=\left(\frac{4\pi}{3}\right)^{1/3}(c_1+c_2+c_3+c_{12}+c_{13}+c_{23})\, ,
\end{equation}
\begin{equation}
 \eta_1=\frac{c_1}{c_1+c_2+c_3+c_{12}+c_{13}+c_{23}}\, , 
\end{equation}
\begin{equation}
 \eta_2=\frac{c_2}{c_1+c_2+c_3+c_{12}+c_{13}+c_{23}}\, ,
\end{equation}
\begin{equation}
 \eta_3=\frac{c_3}{c_1+c_2+c_3+c_{12}+c_{13}+c_{23}}\, ,
\end{equation}
\begin{equation}
 \eta_{12}=\frac{c_{12}}{c_1+c_2+c_3+c_{12}+c_{13}+c_{23}}\, ,
\end{equation}
\begin{equation}
 \eta_{13}=\frac{c_{13}}{c_1+c_2+c_3+c_{12}+c_{13}+c_{23}}\, ,
\end{equation}
\begin{equation}
 \eta_{23}=\frac{c_{23}}{c_1+c_2+c_3+c_{12}+c_{13}+c_{23}}\, .
\end{equation}
Let us consider the original perovskite structure represented in Fig.~\ref{fig:perovskite}. The proportions of nuclei $(A_1,Z_1)$, 
$(A_2,Z_2)$, and $(A_3,Z_3)$ are $1/5$, $1/5$, and $3/5$ respectively. By symmetry, we have $c_1=c_2$, or 
equivalently $\eta_1=\eta_2$. 

We shall follow the same approach as for the fluorite structure. 
\begin{itemize}
 \item $Z_2=Z_3=0$
 
 The nuclei $(A_1,Z_1)$ form a sc sublattice, thus leading to 
 \begin{equation}\label{eq:p1}
  c_1 = \left(\frac{3}{4\pi}\right)^{1/3} \left(\frac{1}{5}\right)^{4/3} C_{\rm sc} \, . 
 \end{equation}
\item $Z_3=0$ and $Z_1=Z_2$

The nuclei $(A_1,Z_1)$ and $(A_2,Z_2)$ form a bcc sublattice. We thus find 
 \begin{equation}\label{eq:p2}
  2 c_1 + c_{12} = \left(\frac{3}{4\pi}\right)^{1/3} \left(\frac{2}{5}\right)^{4/3} C_{\rm bcc} \, . 
 \end{equation}
 \item $Z_2=0$ 
 
 The cubic perovskite structure coincides with the sc2 lattice shown in Fig.~\ref{fig1}. 
 The corresponding lattice energy density 
 \begin{equation}
 \mathcal{E}_L = n_N \frac{e^2}{a_{N}}\left( c_1 Z_1^2 + c_3 Z_3^2 +  c_{13} Z_1 Z_3 \right)\, , 
 \end{equation}
 can be directly compared to that given in Ref.~\cite{jog82} (see Table~\ref{tab1}). In particular, 
 the coefficients $c_3$ and $c_{13}$ can be completely determined from the equations: 
 \begin{equation}\label{eq:p3}
  c_1 + c_3 + c_{13} = \left(\frac{3}{4\pi}\right)^{1/3} \left(\frac{4}{5}\right)^{4/3} C_{\rm fcc} 
 \end{equation}
 (the nuclei $(A_1,Z_1)$ and $(A_3,Z_3)$ form a fcc sublattice), 
 \begin{equation}\label{eq:p4}
  \eta_{\rm sc2} = \frac{c_3}{c_1 + c_3 + c_{13}} \, .
 \end{equation}
  
  \item $Z_1=Z_2=Z_3=1$
  
 The lattice energy density reduces to 
\begin{equation}
 \mathcal{E}_L = n_N \frac{e^2}{a_{N}}\left( c_1  + c_2 + c_3 + c_{12} + c_{13} + c_{23}\right)\, , 
\end{equation}
which can be directly compared to the expression obtained in Ref.~\cite{bald92}: 
\begin{equation}
 \mathcal{E}_L = -1.694648\, n_N \frac{e^2}{2 a_{N}}\, .  
\end{equation}
We can thus determine the remaining coefficient $c_{23}$ from the equation 
\begin{equation}
2\left( c_1  + c_2 + c_3 + c_{12} \right)= -1.694648\, , 
\end{equation}
using Eqs.~(\ref{eq:p1}), (\ref{eq:p2}), and (\ref{eq:p3}). 
\end{itemize}

Having determined all the lattice constants of the original perovskite compound, the structure function 
of the binary compounds shown in Fig.~\ref{fig2} can be easily determined from particular cases: 

\begin{itemize}
 \item $Z_1=Z_3$  
  \begin{equation}
   C_{\rm p1}=(2 c_1 + c_3 + c_{12} + c_{13} + c_{23})\left(\frac{4\pi}{3}\right)^{1/3}\, ,
  \end{equation}
  \begin{equation}
   \eta_{\rm p1}=\frac{c_1+c_{13}+c_3}{2 c_1 + c_3 + c_{12} + c_{13} + c_{23}}\, ,
  \end{equation}
  \begin{equation}
   \zeta_{\rm p1}=\frac{c_1}{2 c_1 + c_3 + c_{12} + c_{13} + c_{23}}\, .
  \end{equation}  

 \item $Z_1=Z_2$  
  \begin{equation}
   C_{\rm p2}=C_{\rm p1}\, ,
  \end{equation}
  \begin{equation}
   \eta_{\rm p2}=\frac{2c_1+c_{12}}{2 c_1 + c_3 + c_{12} + c_{13} + c_{23}}\, ,
  \end{equation}
  \begin{equation}
   \zeta_{\rm p2}=\frac{c_3}{2 c_1 + c_3 + c_{12} + c_{13} + c_{23}}\, .
  \end{equation}  
  
 \item $Z_2=Z_3$  
  \begin{equation}
   C_{\rm p3}=C_{\rm p1}\, ,
  \end{equation}
  \begin{equation}
   \eta_{\rm p3}=\zeta_{\rm p1}\, ,
  \end{equation}
  \begin{equation}
   \zeta_{\rm p3}=\frac{c_1 + c_3 + c_{23}}{2 c_1 + c_3 + c_{12} + c_{13} + c_{23}}\, .
  \end{equation}  
\end{itemize}

\begin{acknowledgments}
This work was mainly financially supported by Fonds de la Recherche Scientifique - FNRS (Belgium). Partial support comes also from the COST Action MP1304 ``NewCompStar''. 
The authors thank D. G. Yakovlev and A. A. Kozhberov for discussions. 
\end{acknowledgments}

\end{document}